\documentclass[floats,floatfix,showpacs,amssymb,prd,twocolumn,superscriptaddress,nofootinbib,nolongbibliography,reprint,preprintnumbers]{revtex4-1}

\usepackage{amssymb,amsmath,verbatim,mathtools,needspace,enumitem,etoolbox,graphicx,physics,microtype,afterpage,xspace,tabularx,lmodern,multirow,bm}
\usepackage{textcomp, gensymb}

\usepackage{braket}

\usepackage{relsize}

\usepackage{dcolumn}

\usepackage[normalem]{ulem}
\usepackage{placeins}
\usepackage{comment}
\usepackage[dvipsnames, usenames, table]{xcolor}
\definecolor{linkcolor}{rgb}{0.0,0.3,0.5}
\usepackage[unicode, colorlinks=true, linkcolor=linkcolor, citecolor=linkcolor, filecolor=linkcolor, urlcolor=linkcolor, linktocpage, breaklinks]{hyperref}
\usepackage[all]{hypcap}
\usepackage[T1]{fontenc}
\usepackage[utf8]{inputenc}
\usepackage{mathrsfs}
\hypersetup{colorlinks=true,citecolor=romared,linkcolor=romared,urlcolor=romared}

\definecolor{romared}{RGB}{142,0,28}

\newcommand{\be}{\begin{equation}}
\newcommand{\ee}{\end{equation}}

\def\be{\begin{equation}}
\def\ee{\end{equation}}
\newcommand{\beq}{\begin{eqnarray}}
\newcommand{\eeq}{\end{eqnarray}}

\usepackage{aas_macros}
\usepackage{makecell}
\usepackage{soul}
\usepackage[nolist,nohyperlinks]{acronym}
\usepackage{orcidlink}
\usepackage{lipsum}
\usepackage{booktabs}
\usepackage{bbold}

\acrodef{LSC}[LSC]{LIGO Scientific Collaboration}
\acrodef{BH}{black hole}
\acrodef{NS}{neutron star}
\acrodef{PN}{Post-Newtonian}
\acrodef{BBH}{binary black-hole}
\acrodef{BNS}{binary neutron-star}
\acrodef{NSBH}{neutron-star black-hole}
\acrodef{NR}{numerical relativity}
\acrodef{GW}{gravitational wave}
\acrodef{PSD}{power spectral density}
\acrodef{aLIGO}{Advanced Laser interferometer Gravitational-Wave Observatory}
\acrodef{AZDHP}{aLIGO zero detuned high power density}
\acrodef{GR}{general relativity}
\acrodef{PE}{parameter estimation}
\acrodef{LAL}{LVK algorithm library}
\acrodef{TPI}{tensor-product interpolant}
\acrodef{SVD}{singular value decomposition}
\acrodef{SNR}{signal-to-noise ratio}
\acrodef{ODE}{ordinary differential equation}
\acrodef{PDE}{partial differential equation}
\acrodef{ROM}{reduced order model}
\acrodef{QNM}{quasi-normal mode}
\acrodef{IMR}{inspiral-merger-ringdown}
\acrodef{LVK}{LIGO-Virgo-KAGRA}
\acrodef{SXS}{Simulating eXtreme Spacetimes}

\newcommand{\jhu}{\affiliation{William H. Miller III Department of Physics and Astronomy, Johns Hopkins University, 3400 North Charles Street, Baltimore, Maryland, 21218, USA}}
\newcommand{\ias}{\affiliation{Institute for Advanced Study, Einstein Drive, Princeton, New Jersey, 08540, USA}}

\newcommand{\KITP}{\affiliation{Kavli Institute for Theoretical Physics, University of California Santa Barbara, Kohn Hall, Lagoon Rd, Santa Barbara, California, 93106, USA}}

\graphicspath{{../Plots/}}

\newcommand{\orcid}[1]{\href{https://orcid.org/#1}{\includegraphics[width=10pt]{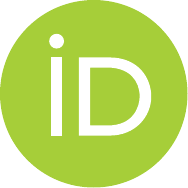}}}
\newcommand*{\rom}[1]{\expandafter\@slowromancap\romannumeral #1@}

\newcommand{\ben}{\begin{enumerate}}
\newcommand{\een}{\end{enumerate}}

\def\be{\begin{equation}}
\def\ee{\end{equation}}
\def\beq{\begin{eqnarray}}
\def\eeq{\end{eqnarray}}

\def\apjl{\rm{ApJ}}
\def\apjs{\rm{ApJS}}
\def\aap{\rm{A\&A}}

\graphicspath{{./Plots/}}

\allowdisplaybreaks

\begin{document}

\pagenumbering{arabic}

\title{Searching for intermediate mass ratio binary black hole mergers in the third observing run of LIGO-Virgo-KAGRA}

\author{Mark Ho-Yeuk Cheung\,\orcid{0000-0002-7767-3428}}
\email{hcheung5@jhu.edu}
\jhu

\author{Digvijay Wadekar\,\orcid{0000-0002-2544-7533}}
\jhu

\author{Ajit Kumar Mehta\,\orcid{0000-0002-7351-6724}}

\author{Tousif Islam\,\orcid{0000-0002-3434-0084}}
\KITP

\author{Javier Roulet\,\orcid{0000-0003-3268-4796}}
\affiliation{TAPIR, Walter Burke Institute for Theoretical Physics, California Institute of Technology, Pasadena, CA 91125, USA}

\author{Emanuele Berti\,\orcid{0000-0003-0751-5130}}
\jhu

\author{Tejaswi Venumadhav\,\orcid{0000-0002-1661-2138}}
\affiliation{Department of Physics, University of California, Santa Barbara, CA 93106, USA}
\affiliation{International Centre for Theoretical Sciences, Tata Institute of Fundamental Research, Bangalore 560089, India}

\author{Barak Zackay\,\orcid{0000-0001-5162-9501}}
\affiliation{Department of Particle Physics and Astrophysics, Weizmann Institute of Science, Israel}

\author{Matias Zaldarriaga\,\orcid{0009-0007-8315-6703}}
\ias

\date{\today}

\begin{abstract}
Intermediate mass ratio inspirals (IMRIs) of binary black holes with mass ratios $10^{-4}\lesssim q \lesssim 0.1$ are astrophysically interesting sources of gravitational waves.
Mergers of intermediate-mass black holes (IMBHs) with stellar-mass black holes would be IMRIs, so their detection can help us probe the formation mechanisms of IMBHs. 
They can also help us perform precise tests of general relativity due to the presence of strong higher-order mode emission.
We perform a search for aligned-spin IMRIs within the data of the two LIGO detectors in the third observing run (O3) of the LIGO-Virgo-KAGRA (LVK) collaboration, including higher modes in the template banks for the first time.
We use the IAS-HM\footnote{\label{foot}The \texttt{IAS-HM} pipeline is publicly available at \url{https://github.com/JayWadekar/gwIAS-HM}} pipeline for our search and construct template banks in the range $1/100 < q<1/18$ using the \texttt{SEOBNRv5HM} waveform model. 
Our banks retain a similar level of effectualness for \texttt{IMRPhenomXHM} and \texttt{BHPTNRSur2dq1e3} waveforms, making our search results relatively robust against waveform systematics.
We show that the sensitivity volume of the search increases by up to $\sim 500\%$ upon inclusion of higher modes.
We do not find any significant candidates with inverse false alarm rate (IFAR) $> 1$ year in the O3 data.
This gives us upper limits on the IMRI merger rate in the local Universe, ranging from $\sim 30$ to $10^3$ Gpc$^{-3}$ yr$^{-1}$ depending on the masses of the black holes in the binary.
These constraints are consistent with rate predictions in the literature. 
Our projections indicate that we would be able to detect IMRIs or constrain some of their proposed formation channels in the fourth (O4) and fifth (O5) observing runs.

\end{abstract}

\preprint{000000}

\maketitle

\section{\label{sec:intro}Introduction}

Gravitational-wave (GW) astronomy became an observational science ten years ago with the first detection of a compact binary coalescence (CBC) event~\cite{LIGOScientific:2016aoc}.
Since then, the LIGO-Virgo-KAGRA (LVK) GW detector network has detected close to a hundred CBCs in the first three detection runs, combining the GWTC~\cite{LIGOScientific:2016dsl, LIGOScientific:2018mvr, LIGOScientific:2020ibl, LIGOScientific:2021usb, KAGRA:2021vkt}, OGC~\cite{NitzCatalog_1-OGC_o1_2018, NitzCatalog_2-OGC_o2_2020,nitz_o3a_3ogc_catalog_2021,nitz20234} and IAS~\cite{Venumadhav:2019tad, Venumadhav:2019lyq, Olsen:2022pin, Mehta:2023zlk, Wadekar:2023gea} event catalogs.
These include astrophysically interesting, even unexpected events, like the first multi-messenger observation of a binary neutron star (BNS) merger~\cite{LIGOScientific:2017ync, LIGOScientific:2017vwq}, a heavy binary black hole (BBH) merger for which the primary black hole (BH) and the remnant BH lie in the pair-instability supernova mass gap~\cite{LIGOScientific:2020iuh}, a BBH with a highly unequal mass ratio of $1:8$ and the secondary in the lower mass gap~\cite{LIGOScientific:2020zkf}, the first BBH with a measurable spin in the progenitors~\cite{LIGOScientific:2020stg}, mergers with one component in the lower mass gap~\cite{LIGOScientific:2020aai, LIGOScientific:2024elc}, and neutron star-BH mergers~\cite{LIGOScientific:2021qlt}.
Upon further analysis of the data, candidate events were found with source-frame primary masses as high as $300 M_\odot$~\cite{Wadekar:2023gea}.
Each of these detections in previously unexplored parameter spaces provided immense scientific value, challenging our understanding of how compact binary mergers are formed, and pushing tests of general relativity (GR) to extreme regimes.

Thus, it is only natural for us to push observational GW astronomy further into unknown territory.
Among the events predicted in theory but yet to be confidently observed are intermediate mass ratio inspirals (IMRIs), i.e., BBH mergers with $10^{-4}\lesssim q \lesssim 0.1$~\cite{Brown:2006pj,Mandel:2007hi,amaro2018detecting}, where $q = m_2/m_1 \leq 1$ is the ratio between the masses of the lighter and heavier BHs.
The merger of an intermediate-mass black hole (IMBH) with a stellar-mass black hole (SBH) would have a mass ratio in this range.

While this is somewhat arbitrary, IMBHs are generally defined to be BHs with a mass $\sim 100 - 10^5M_\odot$. They are considered crucial in understanding BH evolution and the potential seeds from which supermassive black holes (SMBHs) grow~\cite{Greene:2019vlv,Inayoshi:2019fun, haberle2024fast}.
Theoretical IMBH formation channels include direct collapse of gas clouds~\cite{Loeb:1994wv,Bromm:2002hb,Begelman:2006db,Lodato:2006hw,Latif:2022vwc}, collapse of Population III stars~\cite{Fryer:2000my,Bromm:2003vv,Karlsson:2011xx}, or dynamical collision of stars and mergers of SBHs in dense environments, like globular clusters (GCs) or nuclear star clusters (NSCs)~\cite{quinlan1987collapse,quinlan1989dynamical,Miller:2001ez,PortegiesZwart:2002iks,davies2011supermassive,lupi2014constraining,Antonini:2018auk,Rodriguez:2019huv,Kritos:2022non,Kritos:2024upo}.
There have been decades of astronomical searches of IMBHs based on accretion signatures and stellar and gas dynamics~\cite{Greene:2019vlv,haberle2024fast}, but until now, no candidate has been robustly confirmed.
The GW190521 event provides direct evidence of an IMBH, because the remnant BH has mass $\sim 142 M_\odot$~\cite{LIGOScientific:2020iuh}.
Targeted searches for IMBH events were performed by the LVK Collaboration~\cite{LIGOScientific:2017zid,LIGOScientific:2019ysc,LIGOScientific:2021tfm}, as well as with the \text{PyCBC-IMBH}~\cite{Chandra:2021wbw,Chandra:2021xvs} and \text{PyCBC-HM}~\cite{Chandra:2022ixv} pipelines, but no additional confident candidates were found.

\begin{figure*}[t!]
    \centering
    \includegraphics[width=0.9\textwidth]{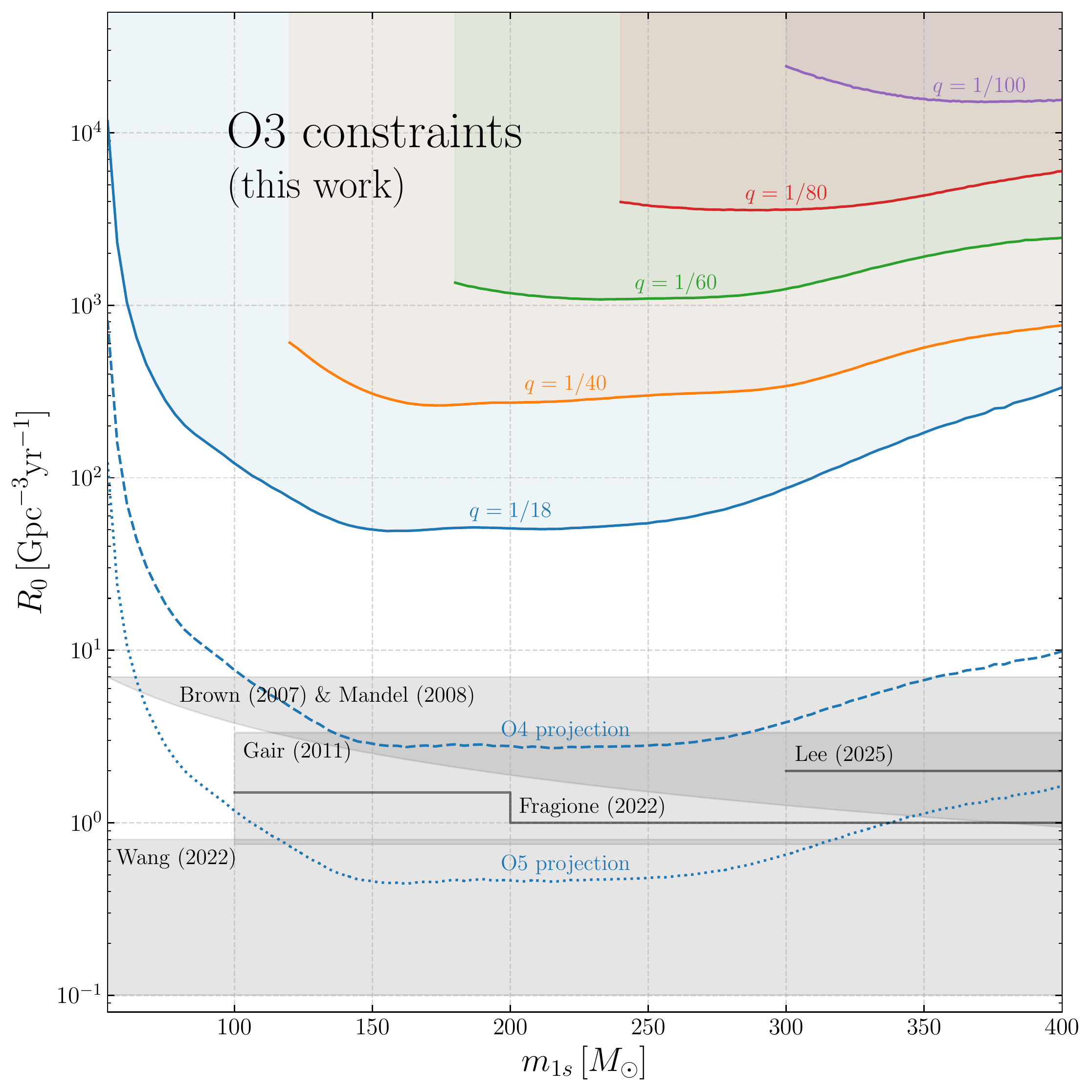}
    \caption{
    The main result of this work: the $90\%$ upper-bound constraints on the local IMRI merger rate density $R_0(m_{1s}, q)$ obtained by our search on the LVK O3 data for mass ratios $1/100 < q < 1/18$ (shaded regions in the upper part of the plot) as a function of the source-frame primary mass $m_{1s}$ of the binary.
    For the precise definition of $R_0(m_1, q)$, see Eqs.~\eqref{eq:lognormals} and~\eqref{eq:r0_m1_q}.
    The curves for lower $q$ are truncated at low masses because we impose $m_{2s} > 3 M_\odot$ for our search.
    The gray shaded regions and lines are the predicted ranges or upper limits of local IMRI rate density in GCs and NSCs by \citet{Brown:2006pj}, \citet{Mandel:2007hi}, \citet{Gair:2010dx}, \citet{Fragione:2022avp}, \citet{Wang:2022unj}, and ~\citet{Lee:2025qbu}, which are obtained by semi-analytical models or numerical simulations (see Sec.~\ref{subsec:rates} for more details).
    For the $q = 1/18$ case, by scaling our O3 constraints assuming that the sensitivity volume-time $\overline{VT} \propto \mathrm{SNR}^3 \, T_{\rm obs}$, we show the projected constraints for O4 (blue dashed line), assuming an observation time of $T_{\rm obs} = 2 \, \mathrm{years}$ at design sensitivity, and O5 (blue dotted lines), assuming $T_{\rm obs} = 3 \, \mathrm{years}$ at A+ sensitivity.
    Note that the projections do not account for the potential difference in the behavior of non-Gaussian noise transients across the observing runs.
    In O4 and O5, we expect to verify or constrain some of the astrophysical models of IMRI formation. 
    }
    \label{fig:rates} 
\end{figure*}

To explain the existence of BHs with masses spanning multiple orders of magnitude from SBHs to SMBHs, it is necessary for BHs to grow their masses in the IMBH range.
One theorized channel of mass growth consists of repeated mergers of a BH with other SBHs, as mentioned above~\cite{quinlan1987collapse,quinlan1989dynamical,Miller:2001ez,PortegiesZwart:2002iks,davies2011supermassive,lupi2014constraining,Antonini:2018auk,Rodriguez:2019huv,Kritos:2022non,Kritos:2024upo}.
At some point during the evolution, the primary BH would be an IMBH of mass $\gtrsim 100 \, M_\odot$, so the merger with an SBH of mass $\lesssim 10 M_\odot$ would be an IMRI with $q \lesssim 0.1$.
These IMRIs could form dynamically in dense environments like GCs and NSCs~\cite{Mandel:2007hi,fragione2018intermediate,Kritos:2022non}, or in the disks of active galactic nuclei (AGNs)~\cite{Derdzinski:2018qzv,Derdzinski:2020wlw}.
Detecting IMRIs and analyzing their properties could allow us to infer their formation channels.
Additionally, the strong contribution of higher multipoles in the gravitational radiation from IMRIs can help us perform a number of tests of GR~\cite{GW190814, GW190412}. 
There has also been recent interest in developing accurate waveform models for IMRIs~\cite{garcia2020multimode,Rin24, Pom23_SEOBNRv5HM, Nag22_IMRI_wf, Isl22_BHPTNRSur1dq1e4, Yoo:2022erv, Lousto:2020tnb, Lousto:2022hoq,Rifat:2019ltp,vandeMeent:2020xgc}.
If the total mass of the IMRI is $\lesssim 400 M_\odot$, the LVK detectors could be sensitive enough to detect the merger~\cite{Brown:2006pj,Mandel:2007hi,amaro2018detecting}.
Thus, a search for IMRIs in current and upcoming LVK data is particularly timely and warranted. 

Searching for GWs sourced by CBCs in the detector data stream is not an easy task.
As these GW signals are faint and the detectors are noisy, many of the observed events have low signal-to-noise ratio (SNR).
While the detector noise is usually approximately Gaussian, sudden and frequent non-Gaussian noise transients, or glitches, complicate the analysis.
Various pipelines have been developed for GW searches, including \texttt{GstLAL}~\cite{messick2017analysis,sachdev2019gstlal,hanna2020fast,cannon2021gstlal}, \texttt{MBTA}~\cite{adams2016low,aubin2021mbta}, \texttt{PyCBC}~\cite{allen2005chi,allen2012findchirp,dal2014implementing,usman2016pycbc,nitz2017detecting,davies2020extending} and \texttt{SPIIR}~\cite{Chu:2020pjv}, which search for GWs by matched filtering the data with waveform templates, and \texttt{cWB}~\cite{klimenko2004wavelet,klimenko2011localization,klimenko2016method}, which searches for transient bursts in the data that are coherent between detectors.
These pipelines have been used by the LVK Collaboration to detect CBCs in the first three observing runs, and the results are compiled into the GWTC catalogs~\cite{LIGOScientific:2018mvr,LIGOScientific:2020ibl,LIGOScientific:2021usb,KAGRA:2021vkt}.
The OGC catalogs were also compiled with additional events from the \texttt{PyCBC} search pipeline~\cite{NitzCatalog_1-OGC_o1_2018, NitzCatalog_2-OGC_o2_2020,nitz_o3a_3ogc_catalog_2021,nitz20234}.
Notably, the OGC searches performed in Refs.~\cite{NitzCatalog_1-OGC_o1_2018, NitzCatalog_2-OGC_o2_2020,nitz_o3a_3ogc_catalog_2021,nitz20234} extend to the IMRI regime and overlap with the parameter space of interest in our paper.

Other than the aforementioned pipelines, the \texttt{IAS} pipeline has also been applied to search for GWs from CBCs in the first three observing runs, identifying events that complement existing catalogs~\cite{Venumadhav:2019tad, Venumadhav:2019lyq, Roulet:2019hzy, Zackay:2019kkv,Olsen:2022pin, Mehta:2023zlk, Wadekar:2023gea,Wadekar:2023kym,Wadekar:2024zdq,Mehta:2025jiq}.
In this work, we use the \texttt{IAS-HM} pipeline~\cite{Wadekar:2023kym,Wadekar:2024zdq} to search for IMRIs in the data from the two LIGO detectors in the third observing run (O3a and O3b) of LVK.
We will focus on IMRIs with $1/100 < q < 1/18$.
Our pipeline is based on matched filtering, which requires building banks of templates and filtering the data with them.
The pipeline constructs the template banks, preprocesses the data, performs the matched filtering, corrects for the drift of the PSD in time, collects coherent triggers, and ranks the collected candidates to determine their false alarm rate (FAR).

Crucially, other than the usually dominant $\ell m = 22$ spherical harmonic mode waveform, the \texttt{IAS-HM} pipeline also includes the $\ell m = 33$ and $44$ higher-order modes (HMs)~\cite{Wadekar:2023kym,Wadekar:2024zdq}. 
{The pipeline uses the \emph{mode-by-mode filtering} approach, which significantly reduces the computational cost of the HM search~\cite{Wadekar:2023kym}. 
Furthermore, it includes a marginalized detection statistic, which allows for efficient marginalization over the extra degrees of freedom corresponding to HMs and increases the pipeline volume sensitivity~\cite{Wadekar:2024zdq, Mehta:2025jiq}. 
These HMs have not been previously included in IMRI searches but they are especially important for IMRIs, as the relative amplitude of HMs increases with the mass asymmetry of the binary.
We search over the public O3 data strain of the two LIGO detectors but find that none of the candidates in the results of our search are significant enough to claim confident detection. 
This allows us to put upper bounds on the rates of IMRIs, as shown in Fig.~\ref{fig:rates}, which is the main result of this work.

The methodology and results of our work are explained in the subsequent sections.
In Sec.~\ref{sec:pipeline}, we explain the details of the \texttt{IAS-HM} pipeline.
In Sec.~\ref{sec:candidates}, we present the IMRI candidates found by our pipeline.
In Sec.~\ref{sec:injection}, we perform an injection-recovery test to determine the sensitivity of our pipeline and constrain the rates of IMRIs in the local Universe.
In Appendix~\ref{app:priors}, we list all of the priors distributions that we used in our work.
In Appendix~\ref{app:amp_cosines}, we discuss the effects of the amplitude of the template banks on the sensitivity of the search pipeline.
In Appendix~\ref{app:22_eff}, we test the effectualness of our template bank if we use only the $\ell m = 22$ mode in both the template and waveforms for testing.
In Appendix~\ref{app:wf}, we discuss  systematics in the current waveform models in the IMRI regime.
In Appendix~\ref{app:weight_banks}, we estimate the fraction of events we expect to detect for each bank and convert the bank-specific false alarm rate to the one combined over all banks.
In Appendix~\ref{app:lvk_cands}, we list the candidates we found that correspond to events already presented in the GWTC-3 catalog.
In Appendix~\ref{app:ifar_cum}, we show that the distribution of the significance of our candidates is statistically consistent with noise.

\section{\label{sec:pipeline}Search Pipeline}

In this section, we describe the \texttt{IAS-HM} search pipeline as implemented for our IMRI GW search.
Most parts of the pipeline are the same as described in Refs.~\cite{Venumadhav:2019tad, Roulet:2019hzy, Zackay:2019kkv, Wadekar:2023kym, Wadekar:2024zdq}.

In brief, after deciding the boundaries of the parameter space that our template banks should cover, we generate a large number of waveform samples in this region by drawing from a prior distribution.
We then simulate the $h_{22}(f)$, $h_{33}(f)$ and $h_{44}(f)$ mode waveforms with the \texttt{SEOBNRv5HM} waveform model~\cite{pompili2023laying,Khalil:2023kep,Mihaylov:2023bkc,Pom23_SEOBNRv5HM}, divide them into banks, and determine an alternative waveform basis for placing templates efficiently with a geometrical placement algorithm.
We construct templates on a uniform grid in the space spanned by this basis, perform matched filtering with them, and collect all triggers that are coincident in both detectors.
These candidate GW events are then ranked by a detection score statistic to determine their significance in terms of their FAR.

\subsection{\label{subsec:template}Template bank}

The \texttt{IAS-HM} pipeline uses matched filtering to search for GWs in the detector data stream.
This requires constructing waveform templates at discrete points in parameter space.
These points must cover the whole parameter space of interest, and should be dense enough such that any GWs potentially in the data must be represented effectually by at least one of the templates.
Our goal is to construct a template bank such that $\gtrsim 90\%$ of our target parameter space will be covered by at least one template with a $\gtrsim 90\%$ match (defined in Eq.~\eqref{eq:match} below).
We do not aim for a higher accuracy, as waveform systematics could be of the same order of magnitude in the IMRI regime. We will come back to this point in Sec.~\ref{subsubsec:effectualness} below.

Our template bank will cover the parameter space region
\begin{align}
    54M_\odot < m_1 &< 400 M_\odot, \nonumber \\
    3M_\odot < m_2 &\equiv q \, m_1, \nonumber \\
    1/100 < q &< 1/18, \nonumber \\
    |\chi_{\rm eff}|<0.9 \quad {\rm and} &\quad |\chi_1|, |\chi_2| < 0.99, \label{eq:template_range}
\end{align}
where $\chi_1$ and $\chi_2$ are the components of the spins of the two progenitor BHs aligned with the orbital angular momentum (i.e., the $z$-components), and the ``effective spin'' $\chi_{\rm eff} = (\chi_1 + q\chi_2)/(1 + q)$ is the sum of the spins weighted by the masses of the BHs.
All masses are measured in the detector frame (redshifted), except when we label them with a subscript $s$ to denote the source-frame mass.
The lower limit of $3M_\odot$ for $m_2$ means that we will only be searching for BBH mergers, excluding neutron stars. This limit translates to a lower limit of $54 M_\odot$ for $m_1$ for the relevant mass ratios.
While BHs with masses $\lesssim 100 M_\odot$ are usually not considered as IMBHs, we extend the mass range down to this regime for completeness.
The upper limit of $q = 1/18$ is chosen because a previous search with the \texttt{IAS-HM} pipeline on the same data was performed for $q > 1/18$~\cite{Wadekar:2023gea,Wadekar:2024zdq}. 
To reduce the dimensionality of the parameter space, we will restrict to merger waveforms for which the two spins are parallel (or anti-parallel) to the orbital angular momentum, i.e., the BBH system is not precessing.
In this case, the $(\ell, m)$ and $(\ell,-m)$ mode waveforms only differ by a factor of $(-1)^\ell$, so it would be trivial to include the negative-$m$ modes in our analysis once we have the positive-$m$ ones.
This allows us to condense the notation, so when we mention the $h_{\ell m}$ mode waveform, it should be understood that the $h_{\ell, -m}$ mode is also included. 

In this work we simulate the waveform samples required for constructing the template bank using the effective one-body \texttt{SEOBNRv5HM} waveform model~\cite{pompili2023laying,Khalil:2023kep,Mihaylov:2023bkc,Pom23_SEOBNRv5HM}.
This waveform model is calibrated to accurate numerical relativity (NR) simulations with $q > 1/20$, and to linear black hole perturbation theory (BHPT) results at $q = 1/1000$.
While the model is not calibrated in the IMRI range we are interested in, it incorporates information from second-order gravitational self-force (2GSF) in the nonspinning modes and radiation-reaction force~\cite{vandeMeent:2023ols}, which is expected to help the model achieve good accuracy in this regime.
Nonetheless, in Sec.~\ref{subsubsec:effectualness} we compare our template bank with other waveform models in the IMRI range to estimate the loss in sensitivity due to systematic errors introduced by waveform modeling.

\subsubsection{Importance of higher modes}

The GWs emitted by a BBH merger can be decomposed in spherical harmonics,
\begin{equation}\label{eq:summed}
    h^+(f) - i h^\times(f) = \sum_{\ell\geq2, \, |m|\leq\ell} {}_{-2}Y_{\ell m}(\iota, \phi_0) h_{\ell m}(f),
\end{equation}
where $h^+(f)$ and $h^\times (f)$ are the two waveform polarizations in the frequency domain, $\iota$ is the inclination angle of the orbital plane of the binary, and $\phi_0$ is the coalescence phase.
The waveform in the detector data stream will be
\begin{equation}\label{eq:polarization}
    h(f) = F_+(\hat{\mathbf{n}}, \psi)h^+(f) + F_\times(\hat{\mathbf{n}}, \psi) h^\times(f),
\end{equation}
where $F_+$ and $F_\times$ are the antenna pattern functions, which depend on the location of the source in the sky relative to the orientation of the detector, i.e., on the right ascension $\theta_\alpha$ and declination $\theta_\delta$, the polarization angle $\psi$, as well as the orientation of the detector, which is a function of time due to the the rotation of the Earth.
We also use $\hat{\mathbf{n}}$ to denote the direction of the source as measured in the frame of the detector.

For BBHs of similar masses ($q \sim 1$), the detected GW $h(f)$ is dominated by $h_{22}$ due to the symmetry of the BBH merger spacetime~\cite{Buonanno:2006ui}.
However, our goal is to search for GWs sourced by IMRIs, i.e., BBHs that have $1/100 < q < 1/18$ according to our definition.
In this case, other waveform modes with higher $\ell m$ will also be important~\cite{Berti:2007fi,Kidder:2007rt,Berti:2007nw}.
For example, the amplitude ratio is $h_{33}/h_{22} \sim (1 - q)/(1+q)$ after a suitable rescaling of the frequency, while $h_{44}/h_{22} \sim 1 - 3 q / (1 + q)^2$~\cite{Garcia-Quiros:2020qpx,Mishra:2016whh}.
Therefore, it is important to include these higher $\ell m$ modes in our template banks to ensure that the templates are effectual to the GWs that we want to detect.
Indeed, it has been shown that when we include these higher modes in our templates, the detection sensitivity volume increases only minimally for $q \sim 1$, but it increases by a factor of $\sim 2$ for $q = 1/4$~\cite{Mehta:2025jiq}.

Our goal is to include all of the $\ell m$ modes necessary to make our template banks effectual, but not too many, because this would introduce a high computational cost or increase false alarm triggers due to the inclusion of more degrees of freedom.
For the IMRI parameter space considered here and defined in Eq.~\eqref{eq:template_range}, we can gauge the importance of including higher $\ell m$ modes by simulating a waveform with and without these modes and calculating its match with the full waveform (including all modes).

We first define a noise-weighted inner product bilinear in the waveforms $h_i(f)$ and $h_j(f)$,
\begin{equation}\label{eq:inner}
    \langle h_i | h_j \rangle = 4 \int^\infty_0 df \dfrac{h_i^*(f)h_j(f)}{S_n(f)},
\end{equation}
where $S_n(f)$ is the frequency-dependent one-sided power-spectral density (PSD) of the detector noise, and an asterisk denotes complex conjugation.
Note that we are required to work in the frequency domain if we use Eq.~\eqref{eq:inner}.
For detector data $d(f)$ and a waveform template $h_{\rm temp}(f)$ (which we assume to be unnormalized), we define the signal-to-noise ratio (SNR) $\rho$ as
\begin{align}
    \rho_\mathbb{h} &\equiv |\langle \mathbb{h} | d\rangle|, \\
    \mathbb{h} &\equiv \dfrac{h_{\rm temp}}{\sqrt{\langle h_{\rm temp} | h_{\rm temp} \rangle}},\label{eq:norm_temp}
\end{align}
where $h_{\rm temp}$ is a waveform template and $\mathbb{h}$ is its normalized version, such that $\langle \mathbb{h} | \mathbb{h}\rangle = 1$.
In our pipeline, the templates are always normalized before performing the search, i.e., we always use $h_{\rm temp} = \mathbb{h}$, but we keep $h_{\rm temp}$ unnormalized in Eq.~\eqref{eq:norm_temp} for completeness.
The subscript in $\rho_\mathbb{h}$ is a reminder that the SNR depends on the template used.
The filter $\langle \mathbb{h} | \cdot\rangle$ is the optimal linear filter for GW detection in the following sense:
if $d = h + n$, with $n$ being Gaussian noise with a PSD $S_n(f)$, the optimal detection statistic for the waveform $h$ is $\rho_\mathbb{h}$ maximized over the time and phase, as long as we choose $h_{\rm temp} = \tilde{A} h$ with $\tilde{A}$ an arbitrary complex constant.
In that case, the expected value of $\rho_{\mathbb{h}}$ takes the optimal value $\rho_{\rm opt} = \sqrt{\langle h | h \rangle}$.
For convenience, we also define the waveform match, which is a normalized version of Eq.~\eqref{eq:inner} maximized over a time difference $\Delta t$ and phase difference $\Delta \phi$ between the waveforms,
\begin{equation}\label{eq:match}
    \langle h_j | h_k \rangle_{\rm norm} \equiv \max_{\Delta t, \Delta \phi} \dfrac{|\langle h_j e^{i(\Delta\phi + f\Delta t)} | h_k \rangle|}{\sqrt{\langle h_j | h_j \rangle\langle h_k | h_k \rangle}},
\end{equation}
so $\langle h_j | h_k \rangle_{\rm norm} \leq 1$, with the maximum value of $1$ achieved when $h_j = h_k$ modulo a difference of $e^{i(\Delta\phi + f\Delta t)}$.
With these definitions, if $d = h + n$ but we use a wrong template $\mathbb{h}^\prime$ to perform matched filtering, the expected value of the SNR output will be $\mathbb{E}[\rho_{\mathbb{h}^\prime}] = \langle h^\prime | h \rangle_{\rm norm} \, \mathbb{E}[\rho_{\mathbb{h}}]$ (because $\mathbb{E}[\langle \mathbb{h}^\prime | n \rangle] = 0$ for any template $\mathbb{h}^\prime$ if the noise is Gaussian with zero mean).
Thus, the normalized match is the appropriate norm for comparing different waveforms in the context of GW detection, because it quantifies the loss in SNR due to using the first waveform as a template to look for a signal described by the second waveform.
The amplitude of GWs depend inversely on the luminosity distance of the source, $h \propto 1/d_L $, so $\rho \propto 1/d_L$ as well.
Thus, imposing a cutoff in $\rho_{\rm cut}$ as a detection threshold corresponds crudely to imposing a cut in luminosity distance $d_{L, \mathrm{cut}}$, i.e., an event is detectable if $d_L \lesssim d_{L,\mathrm{cut}}$.
If BBH merger events are uniformly distributed in comoving volume, the number of events detected would scale as the sensitivity volume $\sim d_{L,\mathrm{cut}}^3$ to a good approximation in the local Universe.
Thus, when a wrong template $\mathbb{h}$ is used to look for a waveform $h$ in the data, the sensitivity volume scales as $\sim \langle \mathbb{h} | h \rangle_{\rm norm}^3$, because $\mathbb{E}[\rho]$ scales as $\sim \langle \mathbb{h} | h \rangle_{\rm norm}$.

\begin{figure*}[t]
    \centering
    \includegraphics[width=0.99\textwidth]{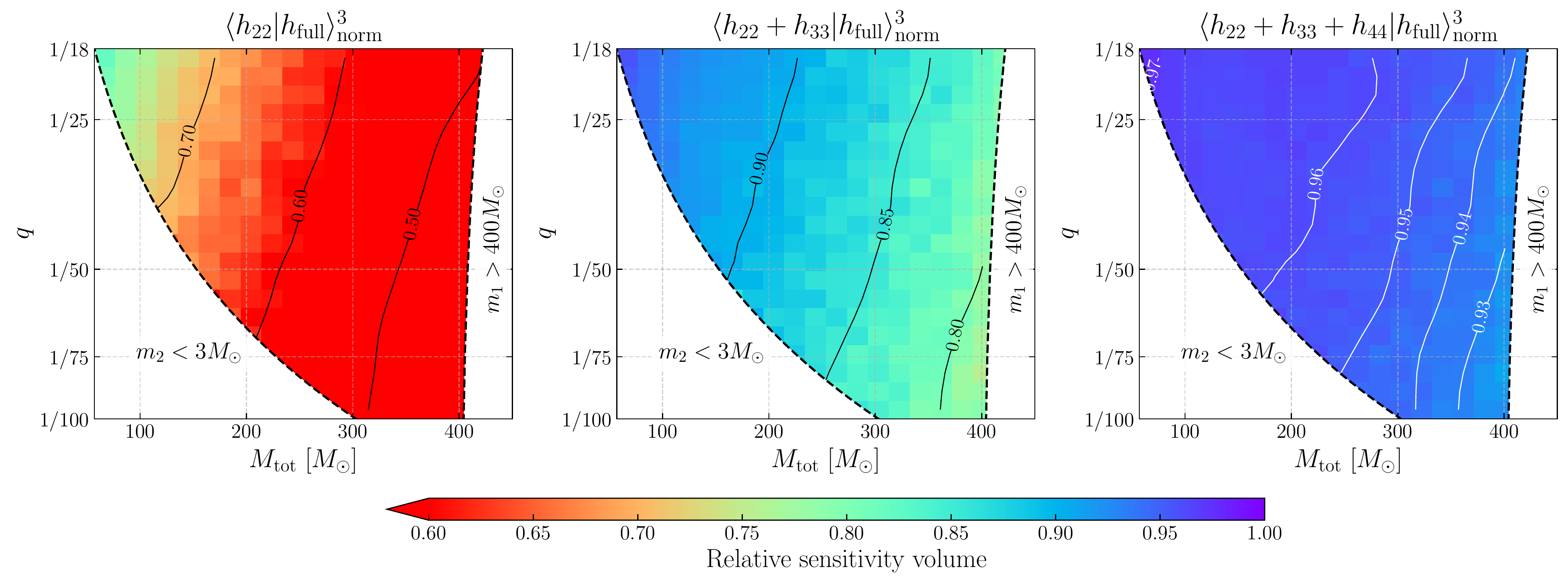}
    \caption{The relative sensitivity volume retained if we only use the $h_{22}$ (left), $h_{22}$ and $h_{33}$ (center), or $h_{22}$, $h_{33}$ and $h_{44}$ waveform modes (right) in our templates, as a function of the detector frame total mass $M_{\rm tot}$ and mass ratio $q = m_2/m_1$.
    The sensitivity volume in this figure is crudely estimated from waveform mismatch. It roughly corresponds to the fraction of events that would be detected by templates with fewer modes versus templates with the full waveform (see Fig.~\ref{fig:VT} for the relative sensitivity volume estimated from an injected population).
    The $h_{33}$ and $h_{44}$ modes are crucial for building an effectual template bank for IMRIs. 
    }
    \label{fig:cosine_missing_lm} 
\end{figure*}

To estimate the sensitivity loss when we only include a few $\ell m$ modes in our templates, we compute the relative sensitivity volume $\langle h^\prime|h_{\rm full}\rangle_{\rm norm}^{3}$, starting with $h^\prime = h_{22}$ and repeating the calculation after adding $h_{33}$ and $h_{44}$.
The relative sensitivity volume gives us an estimate of the fraction of events we expect to detect when our templates are different from the waveforms we want to detect.
The waveforms are simulated with the \texttt{SEOBNRv5HM} model, and the full waveform contains all available modes, i.e., $\ell m = 22, 21, 33, 32, 44, 43, 55$.
In Fig.~\ref{fig:cosine_missing_lm} we show the relative sensitivity volume across the $q$ vs.\ $M_{\rm tot}$ parameter space, where $M_{\rm tot} = m_1 + m_2$ is the detector-frame total mass of the binary.
In each tile of the $M_{\rm tot}$--$q$ plane, we simulate multiple waveforms with an isotropic inclination angle $\iota$, but weighted by the the probability of observing the inclination (which accounts for selection effects, e.g., face-on binaries are louder than edge-on ones~\cite{Sch11}):
\be \label{eq:p_iota}
P(\iota) = \left[ \frac{1}{8}(1+6 \cos^2 \iota +\cos^4 \iota)\right]^{3/2}\, \sin \iota.
\ee

We then take the average relative sensitivity volume by performing a sum over the waveforms associated with that tile.
When going from the left to the right panels, we have added the $33$ mode and the $44$ mode successively, and the sensitivity volume goes from $\lesssim 0.7$ to $\gtrsim 0.9$ for most parts of the parameter space.
This shows that the $33$ and $44$ modes have to be included in the waveform templates to stay within our error budget, but more $\ell m$ modes might not be necessary.
This is a fact we will verify repeatedly in the sections below, where we do a more robust test of the effectualness of our template banks and perform injection-recovery tests.

\subsubsection{Dividing into template banks}

Our procedure for generating template banks follows closely Ref.~\cite{Wadekar:2023kym}, which generates a template bank with only the $\mathbb{h}_{22}$ mode on a grid in parameter space with the geometrical placement algorithm given in~\cite{Roulet:2019hzy}, and then augments the templates by computing also $\mathbb{h}_{33}$ and $\mathbb{h}_{44}$ at each point on the grid.

The first step in using the geometrical placement algorithm is to simulate many waveforms with parameters that adequately cover the target ranges defined in Eq.~\eqref{eq:template_range}.
We draw these samples from a ``waveform samples'' prior, detailed in Appendix~\ref{app:waveform_samples_prior}.
The amplitude of these waveform samples are used to divide the parameter space into different banks, while their phases are used to determine an alternative waveform basis compatible with the geometrical placement approach via singular value decomposition (SVD), as discussed below.
Note that these waveform samples are used to help construct the templates, but they are not the templates themselves.

Our parameter space spans around an order of magnitude in $m_1$, and the behavior of the search is very different when $m_1$ varies within this range.
The density of templates needed decreases drastically when $m_1$ increases, which leads to the look-elsewhere effect being comparatively much larger in the lower mass region.
Also, the characteristic frequency of the templates scales inversely with the mass, meaning that templates with different masses are affected differently by frequency-dependent non-Gaussian noise transients.
This motivates separating the templates into different banks according to the mass, performing the search for each bank separately, and combining the results in the end.
In previous work, the banks were further divided into subbanks, as the duration of the templates within the banks vary significantly.
This however does not happen in the IMRI case, so we do not make use of subbanks.

In practice, rather than dividing the templates according to the mass, we sort the templates into different banks by grouping together templates with a similar amplitude profile $A_{22}(f) = f^{7/6}|h_{22}(f)|$, where the factor of $f^{7/6}$ adjusts for the decay of $|h_{22}(f)|$ in the inspiral phase.
After assigning a bank to each template, we replace the amplitude of all of the waveforms (and later, templates) in the bank by the ``mean'' amplitude of the whole bank, meaning that all of the amplitudes will be set to be the same within the bank.
This ``mean'' amplitude is found by a $k$-means clustering algorithm as implemented in \texttt{sklearn}~\cite{pedregosa2011scikit}.

As long as we use an adequate number of banks, doing this simplifies the search pipeline while introducing minimal loss in the search sensitivity, as the matched-filtering signal-to-noise ratio (SNR) of a GW event is more dependent on the morphology of the phase of $h(f)$~\cite{Roulet:2019hzy}.
The main benefit of using the same amplitude for all templates in a bank is so that when we are correcting for the drift in time of the power spectral density (PSD) of the detector noise, we are only required to compute it once for a bank, because it only depends on the amplitude profile (as a function of frequency) of the signal \cite{Zackay:2019kkv}.
The mean amplitude $A_{22,{\rm mean}}$ and the parameter space covered by each bank are plotted in Fig.~\ref{fig:template_bank}.
As shown in the right panel, constructing banks this way would divide the parameter space into strips in the $M_{\rm tot}$--$\chi_{\rm eff}$ plane.

\begin{figure*}[t]
    \centering
    \includegraphics[width=0.99\textwidth]{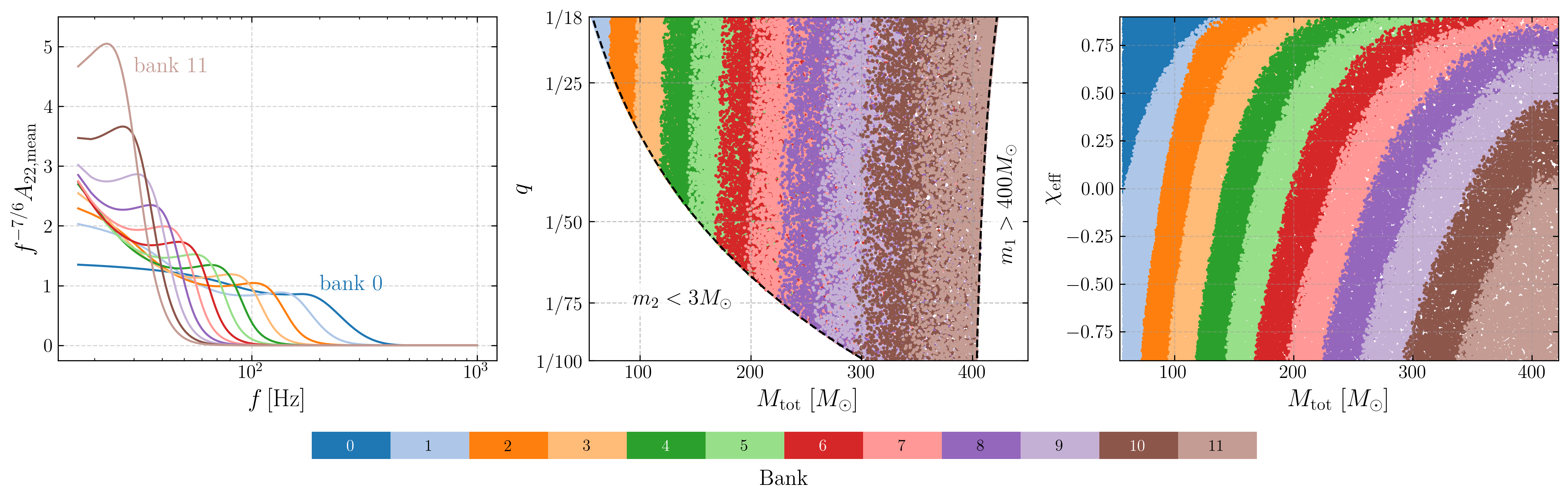}
    \caption{The division of our target parameter space Eq.~\eqref{eq:template_range} into multiple template banks.
    After simulating a large number of waveform samples that cover the parameter space, we divide them into banks by grouping together the samples with similar normalized amplitudes $A_{22} = |h_{22}|/\langle h_{22}|h_{22}\rangle$ by $k$-means clustering.
    Left:
    The mean $A_{22}$ of each bank, which will be used as the one and only amplitude of all $h_{22}$ templates in the bank.
    Center and Right: After dividing into banks, the samples cover different strips of parameter space in the $M_{\rm tot}$-$q$ plane (center) and the $M_{\rm tot}$-$\chi_{\rm eff}$ plane (right).
    Note that these waveform samples are used for generating the templates, but they are not the templates themselves.
    The templates are instead constructed in the space of singular-value-decomposition coefficients, see Sec.~\ref{sec:geometrical_placement} for details.}
    \label{fig:template_bank} 
\end{figure*}

When assigning the waveforms into different banks, we only make use of the $A_{22}$ mode amplitude.
After dividing the banks, we compute the $A_{33,{\rm mean}}$ and $A_{44, {\rm mean}}$ waveform mode amplitudes that correspond to the parameters of the ``mean'' sample picked out by the $k$-means algorithm, and the amplitudes $A_{33}$ and $A_{44}$ of all templates in the same bank are replaced by these mean values.
Again, this is acceptable because the matched-filtering SNR is more sensitive to changes in the phase than the amplitude
(see Appendix~\ref{app:amp_cosines} for an assessment of the sensitivity loss due to this procedure).

\subsubsection{Geometrical placement algorithm}
\label{sec:geometrical_placement}

We have now determined an amplitude to be shared by all templates in a bank.
Again, note that the waveform samples we have simulated are used to facilitate the construction of the template bank, but they are not the templates themselves.
Now, we make use of the phase of the waveform samples.
For a given bank, the geometrical placement algorithm allows us to place a minimal number of templates in the parameter space such that all of the waveform samples are effectually represented by a least one template.
To do so, instead of computing the waveforms on a grid in the conventional $(m_1, m_2, \chi_1, \chi_2)$ parameter space, we need to work in an alternative waveform basis.
This basis is obtained by performing singular value decomposition (SVD) on the phase $\phi_{\ell m}(f) = \arg (h_{\ell m})$ of the waveform samples, using the matched-filtering inner product in~\eqref{eq:inner} as the norm.
Constructing templates on a uniform Cartesian grid in this basis guarantees that the template space is effectually and efficiently covered.
The details of this algorithm can be found in Refs.~\cite{Roulet:2019hzy} and~\cite{Wadekar:2023kym}.
In the end, all of the phases of the simulated waveforms $\arg(h_{22})$, $\arg(h_{33})$ and $\arg(h_{44})$ are represented by sets of coefficients $\{c^{22}_n\}$, $\{c^{33}_n\}$ and $\{c^{44}_n\}$ after projecting onto the SVD basis, with $n \in \{0, 1, 2, \dots\}$ labeling the coefficients of each basis.
Similar to Ref.~\cite{Wadekar:2023kym}, only the first few dimensions in the new basis contribute significantly to the phase, so we only have to keep the first few $c^{\ell m}_n$'s.
In fact, by using only the first three dimensions of the $\ell m = 22$ coefficients $c^{22}_0$, $c^{22}_1$ and $c^{22}_2$, with a random forest regressor we can predict the coefficients $c^{22}_{n\geq3}$ as well as all of the $c^{33}_n$'s and $c^{44}_n$'s.
This is not surprising, as the original $(m_1, m_2, \chi_1, \chi_2)$ parameter space we worked with was low dimensional.
Also, the mapping from the original parameter space to the SVD basis space could be many-to-one due to the $q$-$\chi_{\rm eff}$ degeneracy, which reduces the dimensions needed to represent all of the waveforms.
Moreover, dividing the parameter space into different banks might already have taken into account one of the dominant dimension: see the right panel of Fig.~\ref{fig:template_bank}.
For the regressor, we use the \texttt{RandomForestRegressor} class implemented in \texttt{scikit-learn}~\cite{pedregosa2011scikit}.
We use a uniform grid with spacing $\Delta c_n = 0.3$ spanning the parameter space region of interest, constructing a template on each point of the grid.

\subsubsection{Effectualness of the template bank}\label{subsubsec:effectualness}

\begin{figure*}[t]
    \centering
    \includegraphics[width=0.99\textwidth]{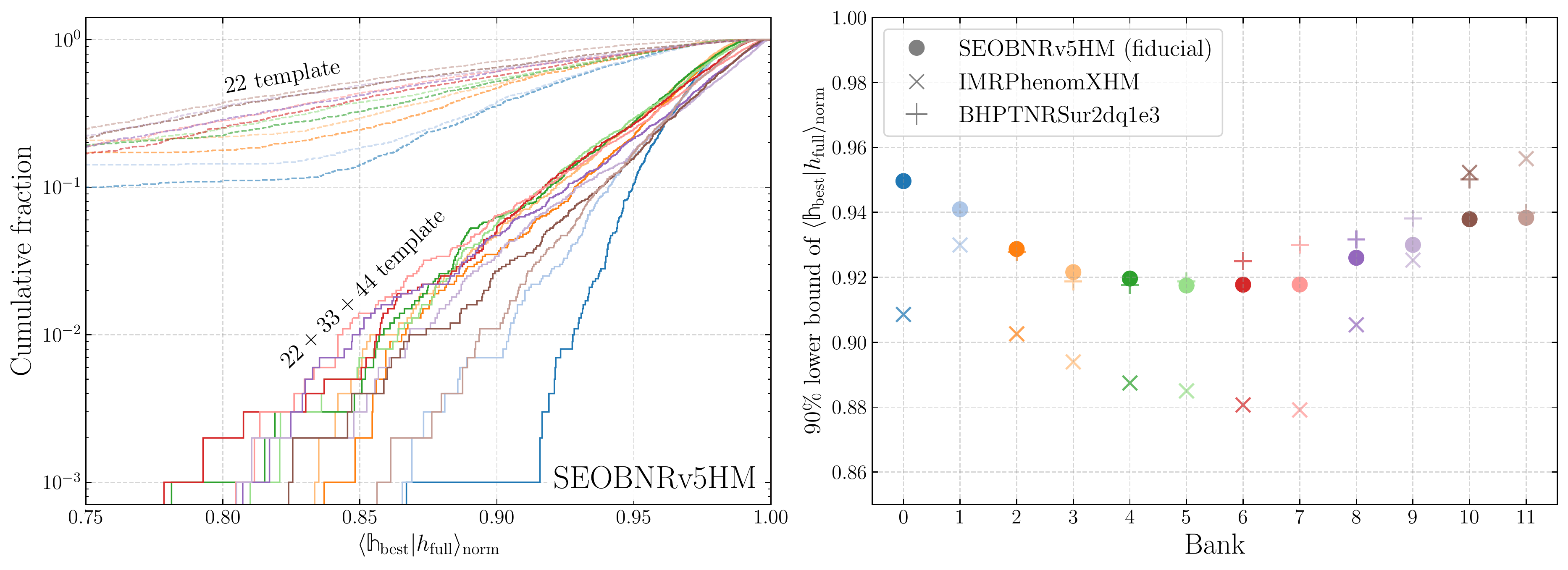}
    \caption{Effectualness of the template bank, tested with a set of waveforms covering the parameter space of each bank, sampled from the prior explained in Appendix~\ref{app:waveform_samples_prior}.
    Left: The cumulative fraction of the match between the sampled waveforms and its corresponding best matching template. 
    The cumulative fraction for each bank is plotted in a different color with solid lines.
    We also compare the same fraction for a template bank constructed with only the $22$ mode (dashed lines), which shows a significant loss in effectualness.
    Right: The $90 \%$ lower bound of the match between the samples and the corresponding best template for each bank, i.e., $90\%$ of the samples within the bank have a template match higher than the labeled points.
    Our template bank is constructed using waveform samples simulated with the \texttt{SEOBNRv5HM} model, but we also test the templates with waveform samples simulated with the \texttt{IMRPhenomXHM} and \texttt{BHPTNRSur2dq1e3} models (see Fig.~\ref{fig:cosines_wf} for a direct mismatch comparison between the different models). Note that the test waveforms used in this figure contain all the waveform modes available in the individual models (see Fig.~\ref{fig:22_eff} for the version of this figure for the 22-only case).
    For banks $0$ and $1$,
    we do not test with the \texttt{BHPTNRSur2dq1e3} model, as the surrogate waveforms are too short for low mass binaries. 
    Overall, we see that our banks retain a similar level of effectualness for three different waveform models, making our search results relatively robust against waveform systematics.}
    \label{fig:effectualness} 
\end{figure*}

Before going ahead with the GW search, we check whether the template bank actually covers the target parameter space with the expected effectualness.
This can be done by checking whether most waveforms lying in the parameter space can be effectually represented by at least one template in the banks.
For this purpose, we could in principle reuse the waveform samples we have simulated for constructing the template bank previously, but this might lead to biases because we are testing the templates with the same waveforms that were used to make them.
Therefore, for testing the effectualness of our templates, we simulate a new set of waveforms with parameters sampled from the same prior, i.e. the waveform samples prior explained in Appendix~\ref{app:waveform_samples_prior}.
We divide these new waveform samples into banks according to the same $k$-means clustering model we trained previously.
Then, we can perform the effectualness test on a bank-by-bank basis, i.e., considering only those waveform samples divided into a particular bank and matching them with the templates of the same bank.
Each waveform sample here is a full waveform (containing all available modes) summed together with an inclination angle $\iota$ drawn from an isotropic distribution weighted by the amplitude modulation due to $\iota$ from Eq.~\eqref{eq:p_iota}, which is the same procedure used to generate the full waveforms in Fig.~\ref{fig:cosine_missing_lm}.
For each waveform sample, we find the template that best matches it and record the match.

In Fig.~\ref{fig:effectualness}, for each bank separately, we plot the cumulative distribution of the match between each waveform sample and its best matching template within its bank.
We also perform the same effectualness test but without including the $33$ and $44$ modes in our templates. 
In that case, the effectualness deteriorates drastically (see the dashed lines in the left panel), which shows again the importance of including higher $\ell m$ modes in our templates.
As shown in the right panel, when we test our templates with the waveform samples simulated with the same waveform model as those used for generating the templates, i.e. the \texttt{SEOBNRv5HM} model, more than $90\%$ of the waveform samples have a match of $\gtrsim 90\%$ with at least one template, which is within our error budget.
For reference, in Appendix~\ref{app:22_eff} we perform the same test, but using the $h_{22}$ waveform samples instead of $h_{\rm full}$ (all modes).

As mentioned previously, waveform models are not calibrated to NR simulations in the IMRI regime, so we might have to worry about systematic errors in our waveform models in this range.
To estimate this error, we perform an additional effectualness test, but simulating waveform samples with the \texttt{IMRPhenomXHM}~\cite{garcia2020multimode} and \texttt{BHPTNRSur2dq1e3}~\cite{Rin24} models and computing the match with our template bank constructed above (which used the \texttt{SEOBNRv5HM} model). 
We choose to compare with the \texttt{IMRPhenomXHM} model because it is widely used and is calibrated against linear BHPT waveforms at $q=1/200$ and $1/1000$~\cite{garcia2020multimode}, and with \texttt{BHPTNRSur2dq1e3} because it bridges between the BHPT and NR regimes by including higher-order self-force effects using non-trivial scalings~\cite{Islam:2023aec,Islam:2023jak}.
While the templates only contain the $\ell m = 22, 33$ and $44$ modes, we will use all available waveform modes for the test waveforms, i.e., the $22, 21, 33, 32, 44, 43, 55$ modes for \texttt{SEOBNRv5HM}, the $22, 21, 33, 32, 44$ modes for \texttt{IMRPhenomXHM}, and the $22, 21, 33, 32, 44, 43$ modes for \texttt{BHPTNRSur2dq1e3}.
As shown in Fig.~\ref{fig:effectualness}, even when the waveform samples come from a different waveform model compared to the templates, we still maintain approximately a $\sim 90\%$ match for $\gtrsim 90\%$ of waveforms.
Note that this might be an overestimation of the waveform systematics, as the true waveform might lie somewhere between the models.
A perceptive reader might notice that the effectualness with \texttt{BHPTNRSur2dq1e3} or \texttt{IMRPhenomXHM} waveforms, in a few cases, outperforms the fiducial \texttt{SEOBNRv5HM} model. 
This is because there are fewer higher $\ell m$ modes for the waveforms simulated from those models, i.e., a fair and comprehensive comparison can only be achieved if we also include the $43$ and $55$ modes in those models. 
In Appendix~\ref{app:wf} we show a direct comparison of the mismatch between different waveform models.

The effectualness test in Fig.~\ref{fig:effectualness} allows us to understand the accuracy of the template bank better than the relative sensitivity volume loss estimated by the mismatch in Fig.~\ref{fig:cosine_missing_lm}. 
Naively, one might expect the cube of the effectualness to be equal to the relative sensitivity volume loss, so that the cube of the circles on the right panel of Fig.~\ref{fig:effectualness} should be approximately equal to the values on the right panel of Fig.~\ref{fig:cosine_missing_lm}.
However, this is not the case, as the cube of the effectualness is $\sim 0.7$, but the relative sensitivity volume is $\sim 0.9$.
The difference can be attributed to multiple reasons.
Firstly, in Fig.~\ref{fig:cosine_missing_lm} each tile is colored according to the mean sensitivity volume for all waveforms that lie within the tile, while in Fig.~\ref{fig:effectualness} we show the $90\%$ lower bound of the match within the whole bank, so we should not naively compare the two numbers.
Secondly, the mismatches in Fig.~\ref{fig:cosine_missing_lm} are computed between two waveforms with the same parameters and different number of HMs, but in Fig.~\ref{fig:effectualness} we compute the effectualness between the waveform and the best matching template.
Finally, the templates are only an approximation of the waveforms in a lower-dimension space spanned by the SVD bases, and they only cover the parameter space coarsely.
All of these reasons contribute to the difference between the results between Fig.~\ref{fig:cosine_missing_lm} and Fig.~\ref{fig:effectualness}.

\subsection{Search} \label{subsec:search}

With a template bank in hand, we can go ahead and perform the GW search in the detector data stream.
We will analyze both the O3a and O3b data streams.
For simplicity, we will only include the data stream from the two LIGO detectors, and we will defer an analysis including Virgo and KAGRA data to future work.

\subsubsection{Data preprocessing}

While the detector data noise is approximately Gaussian and stationary, non-Gaussian noise artifacts such as loud glitches and power lines could contaminate our search results.
Therefore, we apply the procedure in Ref.~\cite{Venumadhav:2019tad} to clean the data stream before we perform the search.
In brief, we remove stretches of data with excessive power in certain frequency bands, creating holes in the data, and then we inpaint these holes with a filter as explained in Ref.~\cite{Zackay:2019kkv}.
This procedure might remove some astrophysical GW signals that are loud (with SNR $\gtrsim 20$), but that is not a concern; the O3 data of LIGO--Virgo--KAGRA have already been analyzed extensively, and loud triggers would have already been picked up by previous searches, even if they did not focus on the IMRI regime.

\subsubsection{Detection statistic}

The remaining parts of the search pipeline will be dedicated to searching for and ranking potential GW event candidates.
The procedures discussed in this section follow closely Sec.~\MakeUppercase{\romannumeral 3} of Ref.~\cite{Wadekar:2024zdq}.
Let us denote the signal hypothesis as $\mathcal{S}$ and the null, or noise hypothesis
as $\mathcal{N}$. 
Note that the noise in our detector is not Gaussian in general, but it is useful to also define a Gaussian noise hypothesis $(\mathrm{GN})$.
The Neyman-Pearson lemma~\cite{neyman1933ix} states that the optimal statistic for distinguishing between two hypotheses is the ratio between their evidences,
\begin{equation}
    \exp\left( \frac{\rho_{\rm score}^2}{2}\right) \equiv \dfrac{P(d|\mathcal{S})}{P(d|\mathcal{N})} = \dfrac{P(d|\mathcal{S})}{P(d|\mathrm{GN})}\dfrac{P(d|\mathrm{GN})}{P(d|\mathcal{N})}, \label{eq:neyman-pearson}
\end{equation}
where $d$ is a stretch of detector data.
With the help of the Gaussian noise hypothesis $\mathrm{GN}$, we have factorized the detection statistic into two parts.
The first part is just the usual evidence ratio in Gaussian noise,
\begin{widetext}
\begin{align}
    \dfrac{P(d|\mathcal{S})}{P(d|GN)} 
    &\propto \left. \int dP_{\rm prior}(\theta) \exp\left( \sum_{k\in{\rm detectors}}-\dfrac{1}{2}\langle d_k - h_{k}(\theta)| d_k - h_{k}(\theta)\rangle\right) \middle/ \exp\left( \sum_{k\in{\rm detectors}}-\dfrac{1}{2}\langle d_k| d_k\rangle\right) \right. \label{eq:sig_gn_ratio}
    \\& \approx \sum_{\alpha \in {\rm templates}} P(\alpha) \left[\int d\pi(\theta_{\rm ext}) \int d\Pi (R_{\ell m}|\alpha) \exp\left( \sum_{k} {\rm Re}\left( \langle h_{\alpha,k}(R_{\ell m})|d_k\rangle\right) - \dfrac{1}{2}\langle h_{\alpha,k}(R_{\ell m})|h_{\alpha,k}(R_{\ell m})\rangle\right)\right]  \label{eq:ext_priors_sum}\\
    & \equiv \sum_{\alpha } P(\alpha) e^{\rho^2_{\rm coherent}/2}, \label{eq:coherent_score}
\end{align}
\end{widetext}
where $\theta \equiv\theta_{\rm in} \cup\theta_{\rm ext}$ are the intrinsic and extrinsic parameters that determine the waveform $h_k(\theta)$ in the $k$th detector.
We use $\alpha$ as a label for each template, so $h_{\alpha,k}$ is the waveform corresponding to the $\alpha$th template, which is characterized by the SVD coefficients $c_n(\alpha)$.
However, the $c_n$'s only give the normalized templates $\mathbb{h}_{22}$, $\mathbb{h}_{33}$ and $\mathbb{h}_{44}$, but not their relative amplitudes.
Therefore, in addition to the label $\alpha$, the physical waveform $h_{\alpha,k} \equiv h_{\alpha,k}(R_{\ell m})$ should also be a function of the mode SNR ratios $R_{\ell m} \in \{R_{33}, R_{44}\}$ between the $\ell m$ and $22$ modes,
\begin{equation}
    R_{\ell m} = \dfrac{\sqrt{\langle h_{\ell m}|h_{\ell m}\rangle }}{\sqrt{\langle h_{22}|h_{22}\rangle }}.
\end{equation}

The prior $P_{\rm prior}$ can be factorized into three parts.
The astrophysical prior $P(\alpha)$ of the $\alpha$th template is equivalent to a prior in the intrinsic parameters $\theta_{\rm in} \in \{m_1, m_2, \chi_1, \chi_2\}$, as the templates are constructed on a grid of SVD coefficients $c_n$'s transformed from the intrinsic parameter space.
The prior $\pi(\theta_{\rm ext})$ takes into account the extrinsic parameters $\theta_{\rm ext} \in \{ \iota, \phi_0, D, \psi, \hat{\mathbf{n}}\}$.
Finally, $\Pi(R_{\ell m}|\alpha)\equiv \Pi(R_{\ell m}|c_n(\alpha))$ is the prior for $R_{\ell m}$ given the $c_n$'s of the template $\alpha$.
While $R_{\ell m}$ is fixed given $\theta_{\rm in}$, the mapping from $\theta_{\rm in}$ to the $c_n$'s is many-to-one due to degeneracies, so we need to make use of the conditional distribution $\Pi(R_{\ell m}|\alpha)$. For example, a single template corresponds to a range of $q, \chi_\mathrm{eff}$ values due to the well-known $q-\chi_\mathrm{eff}$ degeneracy. 
We estimate $\Pi(R_{\ell m}|\alpha)$ by computing the unnormalized waveforms $h_{\ell m}$ and $R_{\ell m}$ at many $\theta_{\rm in}$'s, computing the $c_n$'s for these waveforms, and training a normalizing flow model to approximate $\Pi(R_{\ell m}|c_n(\alpha))$~\cite{Wadekar_inprep}.
The integral over the intrinsic parameters can be approximated by a discrete sum over the templates.
The marginalization over $\theta_{\rm ext}$ is performed by the \texttt{cogwheel} package~\cite{Roulet:2022kot,Islam:2022afg,Roulet:2024hwz}.
The marginalization over $R_{\ell m}$ is performed by Monte Carlo integration.
We define the coherent score $\rho^2_{\rm coherent}$ to be twice the logarithm of the quantity in the square brackets of Eq.~\eqref{eq:coherent_score} to make it an analog of SNR.

The second part of the detection statistic is a correction due to the non-Gaussianity of the noise,
\begin{equation}
    \dfrac{P(d|\mathrm{GN})}{P(d|\mathcal{N})} = \prod_{k\in{\rm detectors}}\dfrac{P(d_k|\mathrm{GN})}{P(d_k|\mathcal{N})}.
\end{equation}
While $P(d_k|\mathrm{GN})$ is simply the denominator of the integrand in Eq.~\eqref{eq:sig_gn_ratio}, $P(d_k|\mathcal{N})$ is intractable, as there is no analytical model of the non-Gaussian detector noise.
Therefore, we work with the following approximation:
\begin{equation}
    \dfrac{P(d_k|\mathrm{GN})}{P(d_k|\mathcal{N})}
      \approx \prod_\alpha \dfrac{P(\rho_k|\mathrm{GN}, \alpha)}{P(\rho_k|\mathcal{N}, \alpha)}, \label{eq:nGcorrmarg}
\end{equation}
with $\rho_k \equiv\rho_k(d_k,\alpha) \approx |\langle h_k(\alpha)|d_k\rangle|$ being the single-detector matched-filtering SNR corresponding to the template $\alpha$.
Note that $\rho_k$ is roughly, but not exactly, equal to $|\langle h_k(\alpha)|d_k\rangle|$ because it is corrected for the PSD drift and holes, as explained in Ref.~\cite{Venumadhav:2019tad}.
In Eq.~\eqref{eq:nGcorrmarg} we are assuming that the templates $h_k(\alpha)$ form a complete basis for representing the data $d_k$.
This allows us to transform $d_k$, which is a multidimensional vector (time series), to the $\rho_k(\alpha)$'s.
In other words, the $\rho_k(\alpha)$'s can be considered to be the coordinates of the data in the space spanned by the template basis, obtained by projection with the $\langle \cdot | \cdot \rangle$ norm.  
This is an approximation, because we have a finite number of templates and they do not form a complete basis.

Computing Eq.~\eqref{eq:nGcorrmarg} requires computing the product over all templates.
However, this can be too expensive to compute practically.
Therefore, we further assume that a single template contributes dominantly to the computation of Eq.~\eqref{eq:neyman-pearson}.
Specifically, we compute an approximate version of the detection score $\rho_{{\rm score},\alpha}$ separately for each $\alpha$, ignoring the summation in Eq.~\eqref{eq:coherent_score} and the product in Eq.~\eqref{eq:nGcorrmarg}, just as in Ref.~\cite{Wadekar:2024zdq}.
Then, we select the ``best'' template $\alpha_{\rm max}$ that gives the highest $\rho_{{\rm score},\alpha}$, and we take this as an approximation of the true overall $\rho_{{\rm score}}$.
This is adequate as long as the other templates do not contribute significantly to Eq.~\eqref{eq:det_score_explicit} below.
This assumption could be relaxed in future work.

Considering again Eq.~\eqref{eq:nGcorrmarg}, we replace the product in $\alpha$ by one particular instance of $\alpha$,
\begin{align}
\left. \dfrac{P(d_k|GN)}{P(d_k|\mathcal{N})} \right|_{\alpha}
     &\approx \dfrac{P(\rho_k|GN,\alpha)}{P(\rho_k|\mathcal{N},\alpha)} \label{eq:nGcorr_ratio_1}\\
     &\propto \exp\left(-\dfrac{|\rho_k|^2}{2} - \log P\big(|\rho_k|^2\,\big|\,\alpha, v, \mathcal{N}\big)\right) \label{eq:nGcorrP}\\
     &\equiv \exp\left(\dfrac{\Delta\rho^2_k(\alpha)}{2}\right), \label{eq:nGcorr}
\end{align}
where $P\big(|\rho_k|^2\,\big|\,\alpha, \mathcal{N}\big)$ is the probability of obtaining a matched-filtering squared SNR of $|\rho_k|^2$ with the template $\alpha$ in the non-Gaussian noise hypothesis when a vetoing criteria $v$ is applied (to be explained later).
In summary, the detection statistic for a particular template $\alpha$ is obtained by multiplying Eqs.~\eqref{eq:coherent_score} and~\eqref{eq:nGcorr}, restricting to one $\alpha$, and taking the logarithm,
\begin{equation}
    \rho^2_{{\rm score}, \alpha} \approx \rho^2_{\rm coherent}(\alpha) - \sum_k\Delta\rho_k^2(\alpha )  + 2\log(P(\alpha)), \label{eq:det_score_explicit}
\end{equation}
where we only keep a particular instance of $\alpha$ in the summation in Eq.~\eqref{eq:coherent_score}, and the final detection statistic is
\begin{equation}
    \rho^2_{\rm score} = \max_\alpha \rho^2_{{\rm score}, \alpha} \, ,
\end{equation}
with the argument of the maximum being $\alpha_{\rm max}$.

\subsubsection{Triggering and candidate collection}

Now that we have defined an optimal detection statistic $\rho^2_{\rm score}$, we can search for GWs by going through the whole GW detector data stream and computing the detection statistic over small time steps, at least in principle.
By defining a tolerance in the false alarm rate (FAR) $r$, we can find GWs in the data by identifying the stretches of data with $\rho^2_{\rm score} > c(r)$, where $c$ is a threshold depending on $r$.
However, this is impractical, because computing $\rho^2_{\rm score}$ over the whole data stream is hopelessly expensive computationally.
In practice, we will use a computationally light approximation to screen out stretches of data with a $\rho^2_{\rm score}$ that could be above threshold.
That is, we will compute an approximation to $\rho^2_{\rm score}$ over the whole data stream, keeping only the candidates over some threshold $c_{\rm approx}$, and computing a more accurate $\rho^2_{\rm score}$ for only these candidates, making sure that the threshold $c_{\rm approx}$ is generous enough to not kill any of the events with the actual score $\rho^2_{\rm score} > c(r)$.

The approximate detection score is explained in Appendix A of Ref.~\cite{Wadekar:2024zdq} and is called $\rho^2_{\rm single{\text-}det}$ (for single detector) and $\rho^2_{\rm multi{\text-}det}$ (for multiple detectors) there.
In brief, we first compute $\rho^2_{\rm single{\text-}det}$, an approximate marginalized coherent score for a single detector, over the whole data stream for each detector separately.
Note that this approximate statistic does not include the non-Gaussian correction $\Delta\rho^2_k$, i.e. we approximate the noise to be Gaussian.
Also, note that this approximate score is computed for each and every template $\alpha$.
Then, we compress the $\rho^2_{\rm single{\text-}det}$ time series by downsampling to a time step of $0.1$\,s, keeping only the maximum within a $0.1$\,s bucket.
We then keep only the time bucket with $\rho^2_{\rm single{\text-}det}$ above a threshold, and we call these ``triggers''.
We further screen the triggers by keeping only the buckets with triggers in both detectors within the light travel time between them, and for each of these triggers we find their corresponding best-fit template $\alpha_{\rm max}$ which gives the maximum $\rho^2_{\rm multi{\text-}det}$.
Here $\rho^2_{\rm multi{\text-}det}$ is another approximation of $\rho^2_{\rm score}$ that takes into account the trigger in both detectors.
Now, we have condensed the whole data stream into a relatively manageable number of candidate triggers, and we can in principle compute the more accurate $\rho^2_{\rm score}$ by Eq.~\eqref{eq:det_score_explicit} using $\alpha_{\rm max}$.
However, before we do that, we can apply a vetoing procedure to further screen the candidates.

\subsubsection{Candidate vetoing}

The triggering and candidate collection explained in the previous paragraphs are performed using the approximate scores $\rho^2_{\rm single{\text-}det}$ and $\rho^2_{\rm multi{\text-}det}$.
As mentioned, these do not include the non-Gaussian correction $\Delta\rho^2_k$, so we would have picked up many candidates triggered by non-Gaussian noise transients, especially for higher mass template banks, where the waveforms are shorter and more similar to short-duration noise glitches.
By computing the full score $\rho^2_{\rm score}$, the non-Gaussian correction will down-weight the glitchy candidates, but it will also penalize higher mass GWs with short duration waveforms.

In fact, when collecting triggers, while we have made use of the matched-filtering SNR, we have not checked whether the template fits the data well.
In other words, we might have collected some glitches as candidates simply because they are loud, even if they do not resemble our templates at all.
This motivates a vetoing procedure for discarding candidates that are not well fitted by the best-fit template $\alpha_{\rm max}$.
Each candidate will undergo a number of vetoing tests, as listed and explained in~\cite{Venumadhav:2019tad}.
These include subtracting the template waveform from the data and checking for excess power in the residual, checking for consistency between the matched-filtering score over multiple frequency bins, and examining whether the candidates lie in bad data segments.
These veto test requirements are represented by the symbol $v$ in Eq.~\eqref{eq:nGcorr_ratio_1}.
This further narrows down the list of candidates, and we can finally compute the full score $\rho^2_{\rm score}$ for each of them.

\subsubsection{Collecting background candidates by time-sliding}

We can now address the determination of $P\big(|\rho_k|^2\,\big|\,\alpha, v, \mathcal{N}\big)$ in Eq.~\eqref{eq:nGcorrP}.
This is actually simplified by making use of the trigger-collecting procedure in the previous paragraph.
That is, if we only care about the significant candidates with $\rho^2_{\rm score} > c(r)$, then
\begin{equation}
    P\big(|\rho_k|^2\,\big|\,\alpha, v, \mathcal{N}\big) = P\big(|\rho_k|^2\,\big|\,\alpha, v, \mathcal{N}, \rho^2_{\rm multi{\text-}det}>c_{\rm approx}\big),
\end{equation}
as long as $c_{\rm approx}$ is chosen generously enough not to kill any significant candidates.
This means that by simulating detector noise (without real GWs) in both detectors, performing the triggering and screening procedure with veto tests $v$ in the previous paragraph to collect these false-positive candidates, and making a histogram in $|\rho^2_k|$, we can estimate the distribution $P\big(|\rho_k|^2\,\big|\,\alpha, v,\mathcal{N}\big)$.

\begin{table*}[htbp]
\centering
\setlength{\tabcolsep}{3pt}
\label{tab:gw_candidates}
\begin{tabular}{*{11}{c}}
\toprule
\multirow{2.5}{*}{Candidate Name} & \multirow{2.5}{*}{$t_{\mathrm{gps}}$ [s]} & \multirow{2.5}{*}{bank} & \multicolumn{4}{c}{Best Fit Template Parameters} & \multicolumn{2}{c}{IFAR [yr]} & \multirow{2.5}{*}{$\rho^2_H$} & \multirow{2.5}{*}{$\rho^2_L$} \\
\cmidrule(lr){4-7} \cmidrule(lr){8-9}
\rule[-3pt]{0pt}{0pt} &  &  & $m_1$ [$M_{\odot}$] & $1/q$ & $s_{1z}$ & $s_{2z}$ & bank & overall &  &  \\
\midrule
\multicolumn{11}{c}{\textbf{O3a}} \\
\addlinespace[0.5em]
\rowcolor{gray!15} 190614\_165038 & 1244566256.14 & 6 & 191.5 & 57.6 & -0.86 & -0.26 & 1.89 & 0.16 & 29.2 & 67.1 \\
 190927\_120913 & 1253621371.60 & 0 & 93.5 & 22.2 & 0.67 & 0.40 & 0.91 & 0.14 & 33.3 & 46.7 \\
\rowcolor{gray!15} 190920\_114731 & 1253015269.66 & 3 & 177.0 & 20.3 & 0.68 & 0.63 & 0.68 & 0.12 & 32.0 & 41.5 \\
 190707\_083226 & 1246523564.95 & 4 & 129.5 & 25.9 & -0.95 & 0.49 & 0.88 & 0.11 & 39.6 & 41.1 \\
\rowcolor{gray!15} 190830\_191935 & 1251227993.87 & 3 & 154.2 & 20.0 & 0.76 & -0.11 & 0.56 & 0.10 & 33.1 & 39.9 \\
 190425\_133124 & 1240234302.59 & 8 & 306.1 & 84.3 & 0.37 & 0.00 & 3.95 & 0.08 & 33.4 & 47.8 \\
\rowcolor{gray!15} 190726\_111313 & 1248174811.09 & 3 & 157.6 & 51.2 & 0.62 & -0.85 & 0.43 & 0.08 & 32.6 & 49.5 \\
 190604\_103812 & 1243679910.94 & 5 & 243.2 & 19.8 & 0.66 & 0.33 & 0.63 & 0.06 & 28.6 & 39.5 \\
\rowcolor{gray!15} 190923\_094358 & 1253267056.87 & 4 & 138.3 & 25.7 & -0.95 & 0.57 & 0.33 & 0.04 & 28.6 & 36.6 \\
 190615\_122524 & 1244636742.45 & 2 & 147.5 & 45.3 & 0.91 & -0.82 & 0.21 & 0.04 & 31.5 & 40.7 \\
\addlinespace[1.0em]
\multicolumn{11}{c}{\textbf{O3b}} \\
\addlinespace[0.5em]
\rowcolor{gray!15} 200305\_225311 & 1267484009.67 & 0 & 71.0 & 19.2 & 0.44 & 0.67 & 1.13 & 0.15 & 52.2 & 38.2 \\
 191202\_203825 & 1259354323.29 & 7 & 237.2 & 64.8 & -0.03 & -0.11 & 1.87 & 0.09 & 25.6 & 65.2 \\
\rowcolor{gray!15} 200210\_173521 & 1265391339.17 & 2 & 141.4 & 20.5 & 0.90 & -0.10 & 0.43 & 0.08 & 41.3 & 31.7 \\
 191230\_153726 & 1261755464.06 & 2 & 149.8 & 21.6 & 0.84 & -0.64 & 0.25 & 0.05 & 37.2 & 35.2 \\
\rowcolor{gray!15} 191205\_153108 & 1259595086.71 & 1 & 95.1 & 26.4 & 0.53 & -0.54 & 0.53 & 0.05 & 27.2 & 47.1 \\
 200129\_114245 & 1264333383.08 & 4 & 170.6 & 28.7 & 0.18 & 0.48 & 0.31 & 0.04 & 28.5 & 42.8 \\
\rowcolor{gray!15} 191209\_191159 & 1259953937.32 & 7 & 288.7 & 53.9 & 0.53 & -0.70 & 0.86 & 0.03 & 35.9 & 38.1 \\
 200113\_205143 & 1262983921.91 & 3 & 159.5 & 18.7 & 0.74 & 0.04 & 0.20 & 0.03 & 37.6 & 46.2 \\
\rowcolor{gray!15} 200108\_104227 & 1262515365.71 & 5 & 172.6 & 27.7 & -0.07 & -0.28 & 0.27 & 0.03 & 25.2 & 48.5 \\
 191127\_114537 & 1258890355.12 & 7 & 274.7 & 51.3 & 0.44 & 0.70 & 0.66 & 0.02 & 23.3 & 53.4 \\
\bottomrule
\end{tabular}
\caption{The top ten candidates with the best overall IFAR in the O3a and O3b data recovered by our pipeline.
The columns are the names of the candidates (their \texttt{YYMMDD\_HHMMSS} time stamp), the GPS time $t_{\rm gps}$ of the trigger at the LIGO Hanford detector, the bank of the template that gave the maximum score, the approximate intrinsic parameters corresponding to that template, the IFAR, both estimated specific to the bank and combined over all banks (see Appendix~\ref{app:weight_banks}), and the squared matched-filtering SNRs $\rho^2_H$ and $\rho^2_L$ of the triggers in the LIGO Hanford and Livingston detectors.
The candidates are listed in order of their significance, i.e., the overall IFAR.
None of these candidates have an overall IFAR $> 1$ year.
Note that this list does not include our search candidates corresponding to events already detected by other pipelines with more equal mass ratios; those are reported in Table~\ref{tab:gw_candidates_lvk} of Appendix~\ref{app:lvk_cands}.
}
\end{table*}

In practice, there is no simple model for simulating non-Gaussian noise in detectors, so we will produce the required noise realizations by ``time-sliding'' the real data stream of detectors.
The idea behind the time-sliding method is that real GW events will induce triggers in different detectors almost at the same time, within the light travel time between the detectors.
Therefore, if we offset (slide) the time between the detectors by an amount larger than the light travel time before feeding the data to our detection pipeline, the candidates obtained will only consist of coincidental noise triggers.
By collecting the candidates over many realizations of this time-sliding procedure, we can estimate $P\big(|\rho_k|^2\,\big|\,\alpha, v,\mathcal{N}\big)$ accurately.
In this work, we will collect candidates on $N_{\rm slide} = 2000$ different realizations of the time slides for this purpose.
We will call these ``background'' candidates, and we will call the actual candidates from the real detector data without time-sliding the ``foreground'' candidates.

\subsection{False alarm rate calculation}

By time-sliding and collecting background candidates, we can also estimate the FAR of a foreground candidate \cite{usman2016pycbc}.
First, we rank all the background candidates from the $N_{\rm slide}$ time slides according to the optimal detection statistic $\rho^2_{\rm score}$.
Then, for a particular foreground candidate, we can compute its $\rho^2_{\rm score}$ and deduce its rank $n_{\rm rank}$ if it were to be inserted among the background candidates
(i.e., combining the triggers from the $N_{\rm slide}$ time slides, there are $\sim n_{\rm rank}$ background candidates with a higher score).
In other words, a false alarm with a score greater than or equal to that of the candidate in question occurs $\sim n_{\rm rank}$ times per $N_{\rm slide}\times T_{\rm obs}$.
The FAR of the candidate is then $n_{\rm rank}/N_{\rm slide} T_{\rm obs}$.

In practice, we perform the triggering and candidate collection procedures described in this section for each template bank separately.
The background candidates collected from time-sliding will be different for each bank, meaning that the FAR computed is bank-specific, i.e., it is the rate of false alarm candidates when the search is performed with only a specific bank.
In Appendix~\ref{app:weight_banks} we explain how we estimate the fraction of events we expect to detect for each bank, which is useful for converting the bank-specific FAR to the one combined over all the banks.

It is also useful to define the inverse false alarm rate by ${\rm IFAR} \equiv 1/{\rm FAR}$, which is the timescale within which we expect there to be one false alarm with a score greater than or equal to a given score.
A candidate event is more statistically significant if the ${\rm IFAR}$ is higher.

\section{Candidates}\label{sec:candidates}

To search for IMRI GWs, we apply the procedure in the previous section to the O3a and O3b data of the LIGO detectors.
For simplicity we do not include the data from the Virgo and KAGRA detectors for the current search; we defer this to future work.
After preprocessing and discarding bad data, we are left with $\sim 106$ days of data in O3a, and $\sim 96$ days of data in O3b.
We then search for coincident triggers with the template bank we constructed in Sec.~\ref{subsec:template} and rank them according to $\rho_{\rm score}^2$ as discussed in Sec.~\ref{subsec:search}.
We perform the triggering and ranking part of our search separately for O3a and O3b.

We report the ten most significant candidates for each observing run in Table~\ref{tab:gw_candidates}, i.e. those with the highest IFAR.
The most significant candidate is \texttt{190614\_165038} in O3a, with an IFAR of $0.16 \, {\rm years}$, 
but none of the candidates have a high enough IFAR for us to claim a confident detection.
We also found some of the events already reported in GWTC-3, although they are not IMRIs: see Appendix~\ref{app:lvk_cands}.

For a combined $T_{\rm obs}$ of $\sim 202$ days, there is a $\sim 3\%$ probability of finding no candidates with an IFAR $> 0.16 \, {\rm years}$, assuming that there are no real GW signals and that noisy candidates follow a temporal Poisson process.
If, instead, we combined the IFARs across banks by assuming that all the banks weigh the same, i.e., $w_b = 1/12$ for all $b$ in Eq.~\eqref{eq:bank_weights},
then the candidate \texttt{190425\_133124} will have an overall IFAR of $0.33$ years and become the top candidate, and the probability of finding no candidate with a better IFAR is $\sim 19\%$.
This implies that our results are consistent with non-detection of IMRI events. 
A more detailed study is presented in Fig.~\ref{fig:ifar_cum} of Appendix~\ref{app:ifar_cum}.

\section{Pipeline sensitivity and rate constraints}\label{sec:injection}

While we did not find any confident IMRI candidates, the non-detection of such events allows us to constrain their rates in the local Universe.
However, while the absence of detections could be because of low astrophysical rates, our pipeline might also be missing a significant fraction of events due to low sensitivity.
To obtain constraints on the astrophysical rates, we need to quantify the selection effects, or the fraction of events we should expect to detect as a function of the parameters of the GW source.
This can be done by performing an injection-recovery test, i.e.
simulating IMRI GWs, injecting them into detector noise, and seeing if our search pipeline can recover them.

\begin{figure*}[t!]
    \centering
    \includegraphics[width=0.99\textwidth]{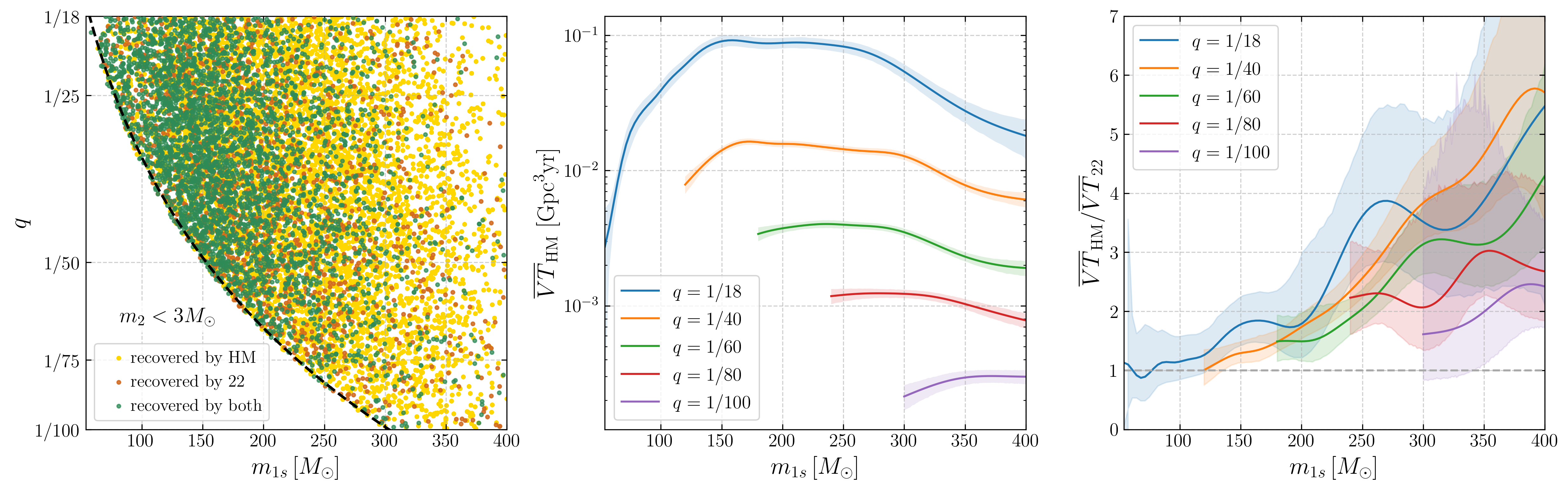}
    \caption{
    Determining the sensitivity of our search pipeline by performing an injection-recovery test.
    Left: The injections recovered by the full higher-mode pipeline and those recovered by the $22$-only pipeline in the $m_{1s}$-$q$ plane, where $m_{1s}$ is the source-frame mass of the primary.
    As expected, the full pipeline recovered more injections.
    Center: The sensitivity volume-time $\overline{VT}$ (defined in Eq.~\eqref{eq:VT_monte_carlo}, using the log-normal in mass astrophysical prior of Eq.~\eqref{eq:lognormals}) of the full higher-mode pipeline for different $q$ at $\mathrm{IFAR} = 1 \, \mathrm{year}$.
    The shaded region is the $90\%$ confidence interval estimated by bootstrapping.
    The curves for more unequal mass ratios are truncated at low masses because $m_{2s} > 3 M_\odot$.
    Right: The ratio between the $\overline{VT}$'s of the full HM and $22$-only pipelines.
    This is the ratio of the number of events expected to be detected between the pipelines.
    The HM pipeline sensivite volume outperforms the $22$-only one by $\sim 50\% - 500$\% at higher masses, depending on $q$. 
    Including higher modes in the template banks thus helps us perform the most sensitive search for aligned-spin IMRIs to date.
    }
    \label{fig:VT} 
\end{figure*}

\subsection{Injection-recovery test}\label{subsec:injection_recovery}

To simulate the waveforms for injection, we draw parameter samples from a prior probability distribution $P_{\rm inj}(\theta)$.
This prior is chosen such that, when compared to the astrophysical prior, we have more injection samples in the parameter ranges where the sensitivity of our pipeline is lower.
This ensures that we have enough injections recovered by the pipeline even in the corners of parameter space where the sensitivity is low, so that we can estimate the selection effect to a good accuracy even in those regions.
Later, the results are reweighted to the astrophysical prior in order to obtain the astrophysical rate constraints.
The probability density function $P_{\rm inj}(\theta)$ used for the injection prior is detailed in Appendix~\ref{app:injection_prior}.
While our templates assumed that the spins of the progenitor BHs are aligned (or anti-aligned) with the orbital angular momentum, we relax this assumption when creating injection waveforms, using the \texttt{SEOBNRv5PHM} waveform model to simulate these waveforms~\cite{pompili2023laying,ramos2023next}, including all available waveform modes.
This allows us to test the sensitivity of our pipeline to fully precessing events.

We then inject the waveforms into noise.
To make sure that the noise is realistic, we inject the simulated waveform into real detector data noise in the O3b run.
We then perform the triggering and candidate collection procedure outlined in Sec.~\ref{subsec:search} and try to recover the injections.
To compute the FAR of each recovered injection, we need a list of background candidates obtained from time-sliding.
We use the same background candidate list that we obtained when performing the real search in Sec.~\ref{sec:pipeline} and~\ref{sec:candidates}, so as to make sure that the conversion from $\rho_{\rm score}^2$ to the FAR for our injection-recovery campaign is the same as the real search.
After imposing a threshold $\tau_{\rm thresh}$ in IFAR, i.e. ${\rm IFAR} > \tau_{\rm thresh}$, we can collect all of the triggers above this threshold and deem them successfully recovered.
In the left panel of Fig.~\ref{fig:VT}, we show the injections that we have recovered with our full HM pipeline, as well as a pipeline using only the $22$ mode templates.
As expected, the HM pipeline recovered more injections.

The number of events we expect to detect within an observation time $T_{\rm obs}$ is
\begin{multline} \label{eq:ndet_int}
    N_{\rm det}[P_{\rm astro}, \tau_{\rm thresh}]
    \\ = R_0 T_{\rm obs}\int dz \, d\theta\, f(z)\dfrac{1}{1+z}\dfrac{dV_c}{dz} P_{\rm astro}(\theta)P_{\rm rec}(\theta, z|\tau_{\rm thresh}) ,
\end{multline}
where $R_0$ is the local ($z = 0$) merger rate density (with units ${\rm volume}^{-1} \times {\rm time}^{-1}$); $P_{\rm rec}(\theta, z | \tau_{\rm thresh})$ is the probability of recovering an injection (i.e. ${\rm IFAR} > \tau_{\rm thresh}$) with parameters $\theta$ and redshift $z$;
$f(z)$ is the evolution of the merger rate density across redshift; and $V_c$ is the comoving volume.
We wrote $N_{\rm det} \equiv N_{\rm det}[P_{\rm astro}, \tau_{\rm thresh}]$ as a reminder that $N_{\rm det}$ depends on the shape of the astrophysical prior and on the IFAR threshold.
We have assumed that the local merger rate density and the astrophysical prior do not depend on time, and also that $P_{\rm astro}(\theta)$ does not evolve with redshift.
In our work, we will assume $f(z) = 1$, i.e. the merger rate does not evolve with redshift, but we will keep this term in our equations for completeness.

\subsection{Sensitivity estimate}\label{subsec:VT}

We can now quantify the sensitivity of our detection pipeline.
We define the detection sensitivity volume-time
\begin{equation}
    \overline{VT}[P_{\rm astro}, \tau_{\rm thresh}] = \dfrac{N_{\rm det}[P_{\rm astro}, \tau_{\rm thresh}]}{R_0},
\end{equation}
which has (of course) units of ${\rm volume} \times {\rm time}$.
This quantity depends on the shape of the merger rate density as a function of $z$ and the parameters $\theta$, but not on its normalization factor $R_0$.
The $\overline{VT}$ is approximately the volume in the Universe that our pipeline is sensitive to, given the shape of the merger rate density, and multiplied by $T_{\rm obs}$.
As we have performed an injection-recovery campaign, we can estimate $\overline{VT}$ for any $P_{\rm astro}(\theta)$ by performing a Monte Carlo integration on Eq.~\eqref{eq:ndet_int}, 
\begin{multline} \label{eq:VT_monte_carlo}
    \overline{VT}[P_{\rm astro}, \tau_{\rm thresh}] \\
    \approx \dfrac{T_{\rm obs}}{N_{\rm inj}} \sum_{{\rm IFAR}_i > \tau_{\rm thresh}} \dfrac{f(z_i)\dfrac{1}{1+z_i}\dfrac{dV_c}{dz} P_{\rm astro}(\vec{\theta_i})}{P_{\rm inj}(\vec{\theta_i}, z)},
\end{multline}
where the summation is only over those injections that have been recovered by our pipeline with an IFAR above $\tau_{\rm thresh}$, because $P_{\rm rec}(\theta, z)$ is $1$ for the recovered injections and $0$ otherwise.
The injection prior $P_{\rm inj}(\vec{\theta_i}, z)$ appears in the denominator because the injection samples are sampled from it, so we basically reweigh the samples to the astrophysical prior.

In the center panel of Fig.~\ref{fig:VT}, we show the $\overline{VT}_{\rm HM}$ estimated by our injection-recovery campaign for the full HM pipeline.
The $\overline{VT}_{\rm HM}$ reduces as $q$ becomes lower because the events becomes less loud.
We also estimate the $\overline{VT}_{\rm 22}$ for our $22$-only pipeline, and we plot the ratio $\overline{VT}_{\rm HM}/\overline{VT}_{\rm 22}$ in the right panel.
This is the final and most powerful confirmation that HMs will improve the search sensitivity for IMRIs overall.
If we did not include the HMs, we could lose a factor of a few in sensitivity volume.

\subsection{Constraints on the IMRI rates}\label{subsec:rates}

In Sec.~\ref{sec:candidates}, we did not find any significant IMRI candidates with ${\rm IFAR} > 1 \, {\rm year}$. 
The most significant candidate that we found had ${\rm IFAR}_{\rm max} = 0.16 \, {\rm years}$.
Given an astrophysical prior $P_{\rm astro}(\theta)$ for IMRIs, we can constrain the local merger rate density $R_0$ as follows.
We can first estimate the $\overline{VT}$ with Eq.~\eqref{eq:VT_monte_carlo}, summing over all of the recovered injections above the threshold ${\rm IFAR}_{\rm max}$.
Naively, as we did not detect any candidate with ${\rm IFAR} > {\rm IFAR}_{\rm max}$ in the real search, $R_0 \overline{VT} = N_{\rm det} \lesssim 1$.
Therefore,
\begin{equation}\label{eq:r0}
    R_0 \lesssim \dfrac{1}{\overline{VT}[P_{\rm astro}, {\rm IFAR}_{\rm max}]},
\end{equation}
which is a constraint on $R_0$.
More formally, GW detection should be a Poisson process, so the right-hand side of Eq.~\eqref{eq:r0} should be multiplied by a factor $\alpha(p) = -\ln(1-p)$, where $p$ is the desired confidence level for the rate limit.

However, as we have not detected any IMRIs yet, we do not know the shape of $P_{\rm astro}(\theta)$ for IMRIs.
Nonetheless, we can still constrain the merger rate density in different ranges of masses and mass ratios $q$ as follows.
We first define a restricted astrophysical prior centered at $m_{1s}$ and $m_{2s} = qm_{1s}$, the primary and secondary masses in the source frame, with a narrow spread $\sigma = 0.1$,
\begin{multline}\label{eq:lognormals}
    P_{m_{1s}m_{2s}}(m_{1s}^\prime, m_{2s}^\prime) \sim {\rm Lognormal}(\mu=m_{1s}, \sigma^2) \\ \times  {\rm Lognormal}(\mu=m_{2s}, \sigma^2).
\end{multline}
Then, we can constrain the rate ``at $m_{1s}$ and $q$'' as 
\begin{equation}\label{eq:r0_m1_q}
    R_0(m_{1s}, q) \lesssim \dfrac{1}{\overline{VT}[P_{m_{1s}m_{2s}}, {\rm IFAR}_{\rm max}]},
\end{equation}
with $m_{2s} = qm_{1s}$.
The definitions in Eqs.~\eqref{eq:lognormals} and~\eqref{eq:r0_m1_q} are consistent with those used by the LVK Collaboration~\cite{KAGRA:2021vkt} as well as the previous IAS pipeline paper~\cite{Mehta:2025jiq}.
Again, a factor of $\alpha(p)$ should appear on the right-hand side, similar to Eq.~\eqref{eq:r0}.

In Fig.~\ref{fig:rates} we show the constraints in $R_0(m_1, q)$ obtained by our full HM search. This is the main result of our paper.
We used ${\rm IFAR}_{\rm max} = 0.16 \, \mathrm{years}$ in Eq.~\eqref{eq:r0_m1_q}, which is the IFAR of the most significant candidate obtained in Sec.~\ref{sec:candidates}.
In our estimate, we have included the Poisson error (i.e. the $\alpha(p)$ factor), the error in the $\overline{VT}$ Monte Carlo integral (Eq.~\eqref{eq:VT_monte_carlo}), and multiplied the constraint upper bound by a factor of $1/0.9^3 \sim 1.37$ to account for the sensitivity loss due to waveform systematics: see the right panel of Fig.~\ref{fig:effectualness}. 

To the best of our knowledge, the constraints in Fig.~\ref{fig:rates} are consistent with all theoretical predictions in the literature~\cite{Brown:2006pj,Mandel:2007hi,Gair:2010dx,Tanikawa:2020cca,Fragione:2022avp,Wang:2022unj,Lee:2025qbu}.
In \citet{Brown:2006pj} and \citet{Mandel:2007hi}, the local IMRI rate density within core-collapsed globular clusters is estimated to be $R_0 \sim 0.07 \times (300 M_\odot/m_{2s}) \mathrm{Gpc}^{-3}\, \mathrm{yr}^{-1}$ by semi-analytic methods, which gives $1$--$7 \, \mathrm{Gpc}^{-3}\, \mathrm{yr}^{-1}$ in our parameter space range.
In \citet{Gair:2010dx}, the rate density is estimated to be $\sim 0.3\times10^9 \,\mathrm{Gpc}^{-3} / T_{\rm merge}$, where $T_{\rm merge}$ is the coalescence time scale, giving $R_0 \sim 0.75$--$3.3\, \mathrm{Gpc}^{-3}\, \mathrm{yr}^{-1}$.
In \citet{Tanikawa:2020cca}, making use of a binary population synthesis code for population III BHs, the rate density is estimated to be $R_0 \lesssim 0.01 \, \mathrm{Gpc}^{-3}\, \mathrm{yr}^{-1}$.
In \citet{Fragione:2022avp}, by simulating the repeated merger of an IMBH with SBHs in NSCs, the rate density is estimated to be $\sim 1.5 \, \mathrm{Gpc}^{-3}\, \mathrm{yr}^{-1}$ for $100 M_\odot < m_{1s} < 200 M_\odot$ and $R_0 \sim 1 \, \mathrm{Gpc}^{-3}\, \mathrm{yr}^{-1}$ for $200 M_\odot < m_{1s} < 500 M_\odot$.
In \citet{Wang:2022unj}, by performing $N$-body simulations in Population III star clusters, the rate density is estimated to be $R_0 \sim 0.1$--$0.8 \,\mathrm{Gpc}^{-3}\, \mathrm{yr}^{-1}$.
In \citet{Lee:2025qbu}, by performing $N$-body simulations in GCs, the rate density is estimated to be $R_0 \sim 2 \, \mathrm{Gpc}^{-3}\, \mathrm{yr}^{-1}$ (see their Fig.~16).
These predictions are all shown as gray regions or gray lines in Fig.~\ref{fig:rates}, with the exception of \citet{Tanikawa:2020cca}, which lies below the plotting range.

Some of the above works converted the rate density to event detection rates $\Gamma$ in the LVK detectors, for example $\Gamma \sim 1 - 30\, \rm{yr^{-1}}$ in Ref.~\cite{Brown:2006pj} and $0.003 - 1.6 \, \rm{yr^{-1}}$ in Ref.~\cite{Arca-Sedda:2020lso}.
As we did not detect any events in O3, our constraint on the event rate is $\Gamma \lesssim 1 \, \rm{yr^{-1}}$, which seems to rule out the prediction in Ref.~\cite{Brown:2006pj} and touch the upper bound of that in Ref.~\cite{Arca-Sedda:2020lso}.
However, without performing a search, the rate density can only be converted to the detection rate by computing the optimal SNR of IMRI waveforms, imposing a detection threshold in the SNR, and approximating the $\overline{VT}$ by computing the volume in the local Universe within which the IMRIs would be detectable above the threshold.
This is a crude approximation, so we choose to compare our results with the predicted rate density $R_0$ instead of the detection rate $\Gamma$.

While our O3 constraints do not rule out any of the astrophysical models of IMRI formation listed above, we can estimate the projected constraints in the O4 and O5 LVK detection runs.
This can be done by assuming that $\overline{VT} \propto \mathrm{SNR}^3 \, T_{\rm obs}$, and that the rate constraints scale as $\overline{VT}^{-1}$.
Assuming that the two LIGO detectors reach design sensitivity~\cite{KAGRA:2013rdx} in O4 and are observing for $T_{\rm obs} = 2 \, \mathrm{years}$, and A+ sensitivity~\cite{LIGO-T1800042} with $T_{\rm obs} = 3 \, \mathrm{years}$ in O5, we compute the SNR of the IMRIs and use $\overline{VT} \propto \mathrm{SNR}^3 \, T_{\rm obs}$ to scale our O3 constraints, obtaining projected constraints for O4 and O5.
As shown in Fig.~\ref{fig:rates}, the projected constraints could reach the rate densities predicted in the literature, so we could verify or constrain astrophysical models for the formation of IMRIs in the next two observing runs.
In scaling $\overline{VT}$ with the SNR, we implicitly assume that the behavior of non-Gaussian noise transients (glitches) remains the same across the observing runs.
However, the signal consistency algorithm for vetoing glitches will become more effective because the signal will remain in band longer for future observing runs, so the constraints attainable could be better than those shown in Fig.~\ref{fig:rates}.

\section{Conclusions}

In this work, we performed a search for IMRIs with $1/100 < q <1/18$ on the O3a and O3b data of LVK.
We made use of the \texttt{IAS-HM} pipeline to construct a template bank and perform the search.
We did not find any new candidates that are significant, with the most significant candidate having an IFAR of $0.16$ years.
We performed an injection-recovery campaign to estimate the sensitivity of our pipeline and constrain the rate density of IMRIs in the local Universe.

Throughout our work, we found that including HMs like the $\ell m =33$ and $44$ modes is crucial for an IMRI search; if only the dominant $22$ mode were included, we could lose a factor of a few in sensitivity volume.
While we did not find any significant candidates, our work proves that an IMRI search with a significant sensitivity is feasible.
In the near future, with more sensitive detectors, more observing time and potentially better waveform models, we could detect IMRIs confidently, or keep improving the constraints obtained here and rule out astrophysical models of IMRI formation channels.

\begin{acknowledgments}
We thank Alessandra Buonanno, Scott Hughes, Konstantinos Kritos, Jonathan Mushkin and Bangalore Sathyaprakash for useful discussions.
M.H.-Y.C., D.W. and E.B. are supported by NSF Grants No.~PHY-2207502, AST-2307146, PHY-090003 and PHY-20043, by NASA Grant No.~21-ATP21-0010, by the John Templeton Foundation Grant 62840, and by the Simons Foundation.
M.H.-Y.C. and E.B. acknowledge support from the ITA-USA Science and Technology Cooperation program, supported by the Ministry of Foreign Affairs of Italy (MAECI), grant No.~PGR01167.
JR acknowledges support from the Sherman Fairchild Foundation.
BZ acknowledges support from the Israel Science Foundation and NSF-BSF 2207583.
MZ acknowledges support from the National Science Foundation NSF-BSF 2207583 and NSF 2209991 and the Nelson Center for Collaborative Research.
This work has made use of the Advanced Research Computing at Hopkins (ARCH) core facility (\url{https://www.arch.jhu.edu/}), which is supported by the NSF Grant No.~OAC-1920103. 
This research has made use of data, software and/or web tools obtained from the Gravitational Wave Open Science Center (\url{https://www.gw-openscience.org/}), a service of LIGO Laboratory, the LIGO Scientific Collaboration and the Virgo Collaboration. LIGO Laboratory and Advanced LIGO are funded by the United States National Science Foundation (NSF) as well as the Science and Technology Facilities Council (STFC) of the United Kingdom, the Max-Planck-Society (MPS), and the State of Niedersachsen/Germany for support of the construction of Advanced LIGO and construction and operation of the GEO600 detector. Additional support for Advanced LIGO was provided by the Australian Research Council. Virgo is funded, through the European Gravitational Observatory (EGO), by the French Centre National de Recherche Scientifique (CNRS), the Italian Istituto Nazionale di Fisica Nucleare (INFN) and the Dutch Nikhef, with contributions by institutions from Belgium, Germany, Greece, Hungary, Ireland, Japan, Monaco, Poland, Portugal, Spain.
\end{acknowledgments}

\appendix 

\section{Prior distributions}
\label{app:priors}

\subsection{Waveform samples prior}\label{app:waveform_samples_prior}

In this work, we use three different prior distributions.
The first one is the waveform samples prior, from which we draw parameter samples and simulate the waveforms for generating the template bank.
These waveforms are not the template themselves, but they are used for dividing the parameter space into different banks, computing the mean amplitude of each bank, finding with SVD the transformed basis used to generate the templates~\cite{Roulet:2019hzy}, and testing the effectualness of the template banks.
Crucially, the waveforms do not have to be sampled from an astrophysical prior.
We simply have to make sure that the sampled waveforms will adequately populate the target parameter space region of Eq.~\eqref{eq:template_range}, making sure that there will be enough samples in each bank.
Thus, for this purpose, we choose a probability density function that could be factorized as follows:
\begin{align}
    P(M_{\rm tot}) &\begin{cases*}
        \mathop\sim^{50\%} M_{\rm tot}^{-1}, \\
        \mathop\sim^{50\%} M_{\rm tot}^{-1/3}, 
    \end{cases*} \\
    P(q) &\sim q^{-1}, \\
    P(\chi_{\rm eff}) &\sim {\rm Uniform}, \\
    P(\chi_1) &\sim {\rm Uniform}.
\end{align}
Half of the samples for the total redshifted mass $M_{\rm tot}$ are sampled from a log-uniform distribution $\sim M_{\rm tot}^{-1}$, and the other half are sampled from a shallower power law distribution $\sim M_{\rm tot}^{-1/3}$ to adequately cover the high-mass banks.
The limits in the spins in Eq.~\eqref{eq:template_range} are respected by first sampling $\chi_{\rm eff}$, and then sampling $\chi_1$ within a slightly restricted range such that the conditions $|\chi_1|, |\chi_2| < 0.99$ are respected.
The templates are only functions of the intrinsic parameters, hence we do not need to sample the extrinsic parameters.

\subsection{Astrophysical Prior for search}\label{app:astro_search_prior}

When performing templated searches, we need to specify astrophysical priors for the intrinsic parameters (i.e., $P(\alpha)$ in Eq.~\eqref{eq:coherent_score}). 
This astrophysical prior is similar to the waveform samples prior, but with a steeper power law in $M_{\rm tot}$:
\begin{align}
    P(M_{\rm tot}) &\sim M_{\rm tot}^{-2}, \\
    P(q) &\sim q^{-1}, \\
    P(\chi_{\rm eff}) &\sim {\rm Uniform}, \\
    P(\chi_1) &\sim {\rm Uniform}.
\end{align}
The $\sim M_{\rm tot}^{-2}$ power law distribution is widely used, and close to the one inferred from the first three observing runs of the LVK collaboration (although usually a mass cut-off at $m_{1s} \sim 80 M_\odot$ is assumed~\cite{LVKO3bpopulation}).
No events within our range of $1/100 < q < 1/18$ have been detected, so we have more freedom to choose the astrophysical prior for $q$.
As our goal is to detect low-$q$ events, we choose a log-uniform distribution such that low $q$ triggers will not be too penalized (compared to a uniform prior in $q$).
The spins are sampled in the same way as the waveform samples prior.

\subsection{Injection Prior}\label{app:injection_prior}

Other than the two priors mentioned above, we also need a prior to generate injection samples for testing our pipeline (see Sec.~\ref{sec:injection}).
This does not need to follow our astrophysical prior because the results of the injection campaign can later be reweighted.
Our goal is therefore to choose an adequate prior such that we have a meaningful chance of recovering some injections even in the corners of the parameter space where the sensitivity of the pipeline is low, e.g. in the high-mass and low-$q$ regimes.
Therefore, we choose the following injection prior, so that more injections will be performed in those regimes:
\begin{align}
    P(M_{\rm tot}) &\sim M_{\rm tot}^{-1}, \\
    P(q) &\sim q^{-4}, \\
    P(\vec{\chi_1}), P(\vec{\chi_2}) &\sim {\text{ Isotropic \& uniform magnitude}}, \\
    P(d_L) &\sim d_L^2, \\
    P(\cos \iota) &\sim \text{Uniform}, \\
    P(\psi) &\sim \text{Uniform}, \\
    P(\theta_{\alpha}, \theta_{\delta}) &\sim \text{Isotropic},
\end{align}
where $d_L$ is the luminosity distance to the binary (which is converted to the redshift $z$ by assuming a $\Lambda$CDM cosmology with Planck Collaboration parameters~\cite{Planck:2018vyg}), $\iota$ is the inclination angle, $\psi$ is the polarization angle of the GWs, and $\theta_{\alpha}$ and $\theta_{\delta}$ are the right ascension and declination of the source in the sky.
We will be injecting precessing waveforms, with the spin directions sampled isotropically and the magnitudes sampled uniformly.
If we sample naively from the above distributions, many samples will have a high value of $d_L$, with most of them undetectable.
One might be inclined to put a hard upper limit on $d_L$, but that would remove the tail in the distribution of recovered injections.
Therefore, we will impose a soft cutoff by multiplying the prior by $P_{\rm det}(\rho_{\rm inj})$, a function of the SNR of the injected waveform $\rho_{\rm inj}$ which approximately follows its detection probability.
Specifically, $P_{\rm det}(\rho_{\rm inj})$ is the probability that the injected waveform will be recovered with an SNR above a threshold of $60$ if it were injected in Gaussian noise, which can be modeled by the survival function of a $\chi^2$-distribution \cite{Roulet:2020}.

\subsection{Restricted astrophysical prior for rate calculation}

\begin{figure}[t!]
    \centering
    \includegraphics[width=0.99\columnwidth]{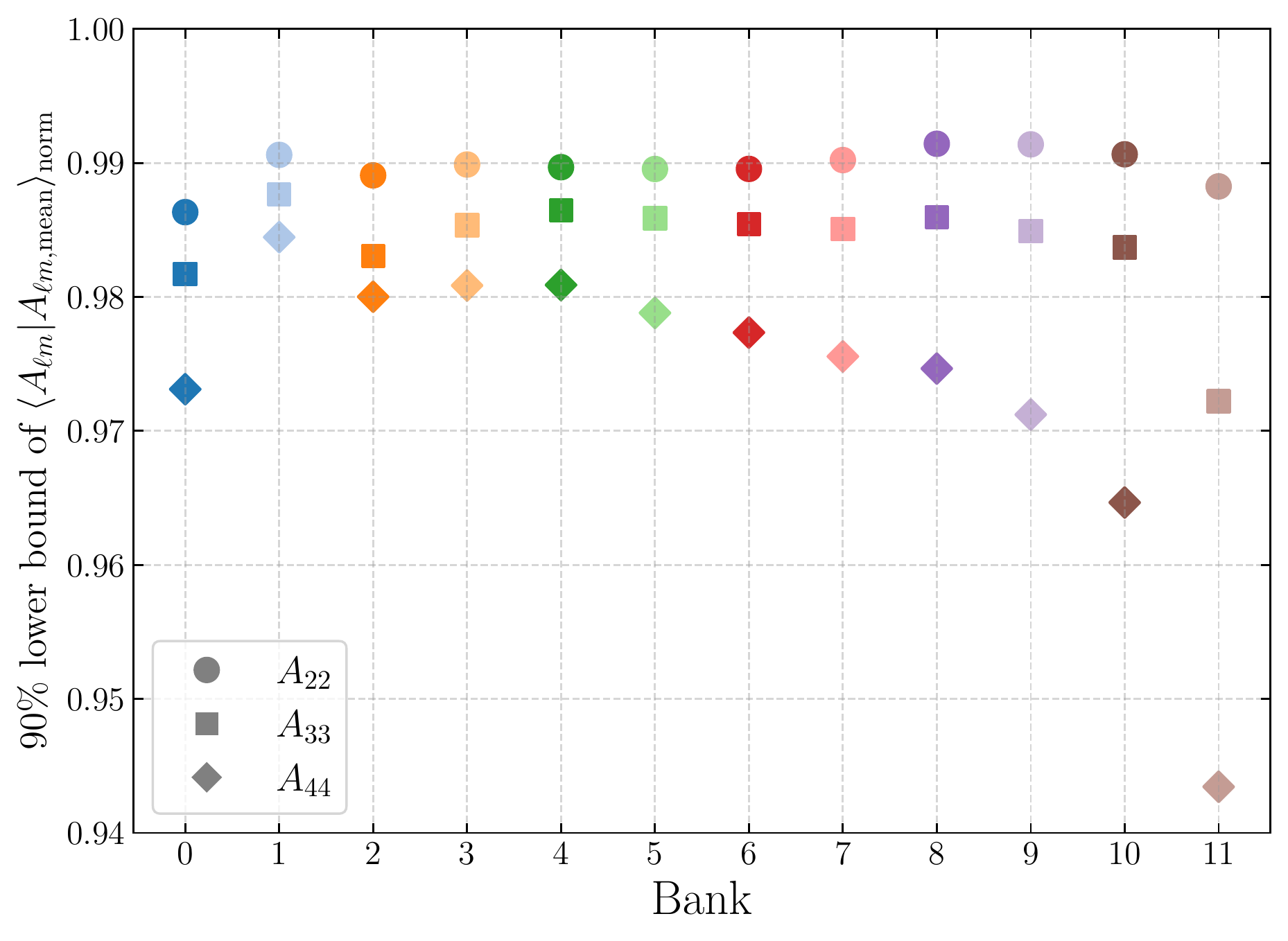}
    \caption{The $90\%$ lower bound of the match between the amplitudes of the waveform samples and the mean bank amplitude, for each bank and each waveform mode ($\ell m = 22, 33, 44$) separately.
    $90\%$ of the samples give a match better than the points shown in this plot.
    As the match is close to unity for all modes and all banks, it is safe to replace them with the mean amplitude of the bank when constructing the templates.
    }
    
    \label{fig:amp_cosines} 
\end{figure}

In the main text, we want to constrain the rates of IMRIs at different masses and mass ratios.
We need to reweight the injection-recovery results to a prior distribution centered at source-frame masses $m_{1s}$ and $m_{2s} = qm_{1s}$, so that we can restrict our attention to the injections around the specified masses and deduce the $\rm VT$ within this restricted range
(see the discussion in Sec.~\ref{sec:injection}).
We use log-normal distributions centered at $m_{1s}$ and $m_{2s}$: 
\begin{align}
    P(m_{1s}) &\sim {\rm Lognormal}(\mu = m_{1s}, \sigma^2=0.1), \\
    P(m_{2s}) &\sim {\rm Lognormal}(\mu = m_{2s}, \sigma^2=0.1), \\
    f(z) &\sim \text{Uniform},
\end{align}
with $m_{2s} = qm_{1s}$, and $f(z)$ the redshift evolution of the merger rate density, as discussed in Sec.~\ref{sec:injection}.
For all the other parameters except $d_L$, we will use the same prior as the injection prior in the previous subsection. 
We will restrict $P(m_{1s})$ and $P(m_{2s})$ to within the limits of the parameter space defined in Eq.~\eqref{eq:template_range}, and make sure that they integrate to unity by adjusting their normalization constants.

\section{Bank amplitudes and sensitivity}\label{app:amp_cosines}

To assess how well the mean amplitude can approximate all amplitudes in a bank (see Sec.~\ref{subsec:template}), we compute $\langle A_{\ell m}|A_{\ell m, {\rm mean}} \rangle_{\rm norm}$, the match between the amplitude of the waveform samples with the mean amplitude of the corresponding bank, using Eq.~\eqref{eq:inner} but ignoring the phase of the waveforms.
As shown in Fig.~\ref{fig:amp_cosines}, when we split all of the templates into $N = 12$ banks (labeling them as bank $0, 1, \dots, 11$), $90\%$ of the waveform samples have an $A_{22}$ amplitude match $> 98\%$ with the mean amplitude of the respective bank, while that of $A_{33}$ and $A_{44}$ are $> 97\%$ and $> 94\%$ respectively,
meaning that replacing all of the template amplitudes in a bank by the mean amplitude will not lead to a significant decrease in sensitivity.
Note that $h_{33}$ and $h_{44}$ are seldom dominant over $h_{22}$ even in the intermediate mass ratio regime, so having a slightly worse match in $h_{44}$ will not degrade the performance of the template bank by a lot.

\section{Effectualness test with $22$-only waveforms}\label{app:22_eff}

\begin{figure}[t!]
    \centering
    \includegraphics[width=0.99\columnwidth]{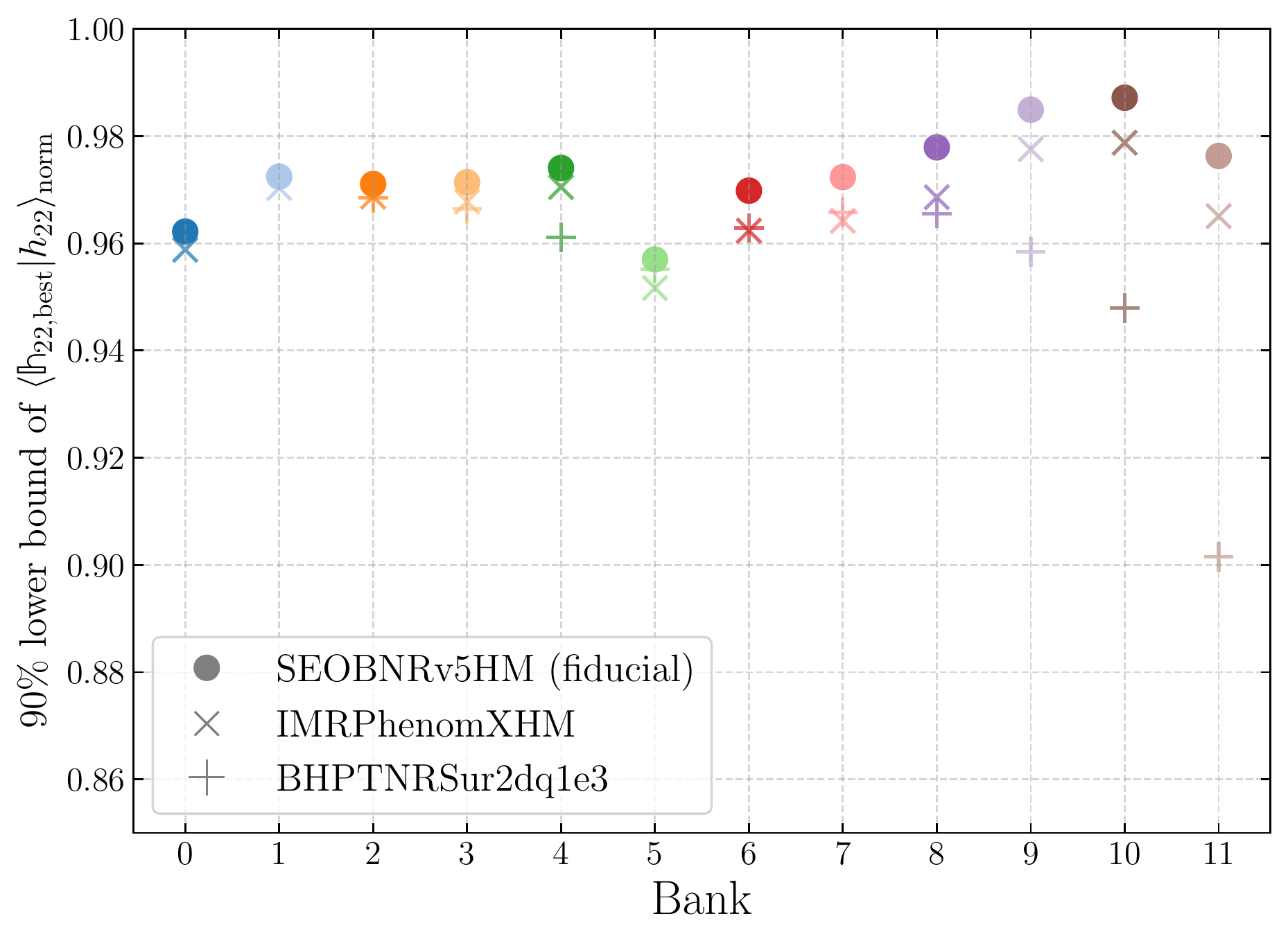}
    \caption{The $90\%$ lower bound of the match between the waveform samples and the corresponding best template, keeping only the $22$ mode in both.
    This plot is similar to the right panel of Fig.~\ref{fig:effectualness}, but the waveforms used there contain all of the available waveform modes of the corresponding waveform model.
    The waveform samples for testing are simulated with the \texttt{SEOBNRv5HM}, \texttt{IMRPhenomXHM} and \texttt{BHPTNRSur2dq1e3} waveform models, but we always use the same template bank constructed with the help of \texttt{SEOBNRv5HM} waveform samples.
    }
    \label{fig:22_eff} 
\end{figure}

\begin{figure*}[t!]
    \centering
    \includegraphics[width=0.8\textwidth]{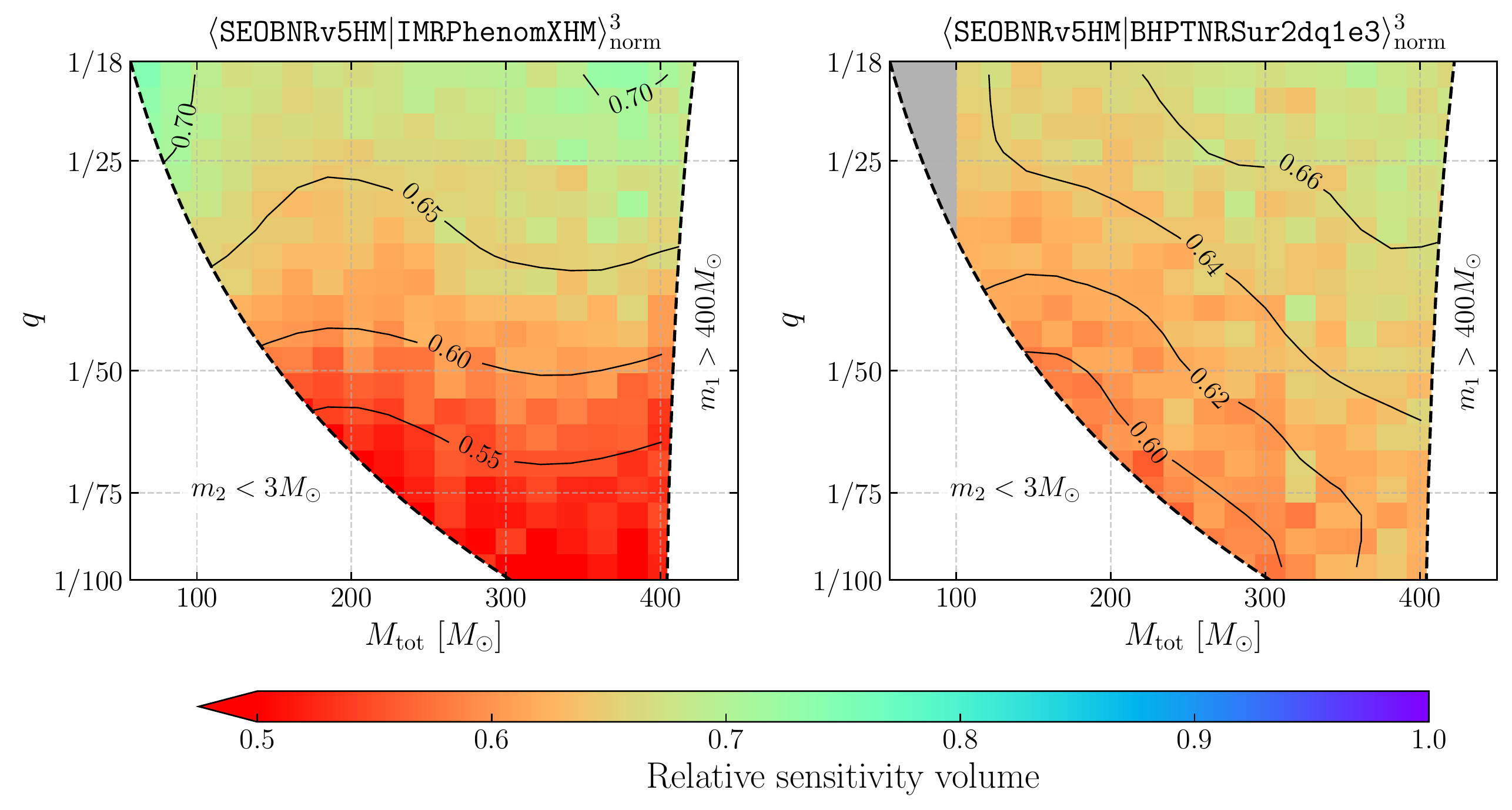}
    \caption{
    The relative sensitivity volume (given by the cube of the mismatch) retained when using the \texttt{IMRPhenomXHM} and \texttt{BHPTNRSur2dq1e3} models instead of \texttt{SEOBNRv5HM}, as an estimation of the effects of waveform systematics on the sensitivity of our pipeline.
    This plot is made following the same procedure as that for Fig.~\ref{fig:cosine_missing_lm}.
    Note that the effectualness shown in Fig.~\ref{fig:effectualness} is better than the matches in this plot because here we are comparing the match between waveforms from different models but with the same exact parameters, while in Fig.~\ref{fig:effectualness} the effectualness is taken to be the match between the waveform and the best matching template across a bank, which do not necessarily have the same parameters.
    We do not compute the sensitivity volume in the gray region of the right panel ($M_{\rm tot} < 100 M_\odot$) because the \texttt{BHPTNRSur2dq1e3} surrogate model is too short to be used for low masses.
    }
    \label{fig:cosines_wf} 
\end{figure*}

In Sec.~\ref{subsubsec:effectualness}, we test our template bank against waveform samples $h_{\rm full}$, which includes all the waveform modes available in the corresponding waveform models.
For completeness, in Fig.~\ref{fig:22_eff} we report the results of the same test, but using only $h_{22}$ in the waveform and only $\mathbb{h}_{22}$ in the template bank.
Naively, we might expect the effectualness to always be better in this case when compared to Fig.~\ref{fig:effectualness}, because in that case there are more HMs in the waveforms than in the templates.
However, when there are HMs in our template bank, the relative amplitudes between the different modes are not fixed. 
When running the full search pipeline, the relative amplitude ratios $R_{\ell m}$ are marginalized over with the help of the prior $\Pi(R_{\ell m}|\alpha)$, see Eq.~\eqref{eq:ext_priors_sum}, but for the effectualness tests in Sec.~\ref{subsubsec:effectualness} we take the values of $R_{\ell m}$ that gives the best match, without any restriction or modulation from a prior.
This allows the HM templates to have more unrestricted degrees of freedom to fit the waveforms better, potentially giving better results when compared to Fig.~\ref{fig:effectualness}.

The results for \texttt{BHPTNRSur2dq1e3} are worse for higher banks in Fig.~\ref{fig:22_eff} because there are more waveforms with negative $\chi_{\rm eff}$ in these banks, and \texttt{SEOBNRv5HM}'s match with it is worse in that corner of parameter space. 

\section{Waveform systematics}\label{app:wf}

As mentioned in Sec.~\ref{subsec:template}, waveform models are not calibrated to NR within our target parameter space defined in Eq.~\eqref{eq:template_range}.
To estimate the sensitivity loss due to systematic errors in the waveforms in this regime, we tested the effectualness of our template banks, which is constructed with \texttt{SEOBNRv5HM}, with waveform samples simulated with two other waveform models, \texttt{IMRPhenomXHM}~\cite{garcia2020multimode} and \texttt{BHPTNRSur2dq1e3}~\cite{Rin24}.
In the right panel of Fig.~\ref{fig:effectualness}, we showed that the loss in effectualness is $\sim 0.1$ for $90\%$ of the waveform samples.
Here, we estimate the loss in sensitivity volume across the parameter space due to waveform systematics with a procedure similar to that for making Fig.~\ref{fig:cosine_missing_lm}, i.e., for each tile in the $M_{\rm tot}$--$q$ plane we simulate waveform samples with different inclinations and sky locations and find the average of the cube of the match within the tile.
For \texttt{BHPTNRSur2dq1e3}, we restrict the primary spin to $|\chi_1| < 0.8$ and the secondary spin to $|\chi_2| = 0$, because the model does not support parameters outside of this range.
As shown in Fig.~\ref{fig:cosines_wf}, the relative sensitivity volume retained is $\sim 0.6$ for both waveform models.
Notice that when comparing across the two panels, at higher $q$'s the match with \texttt{IMRPhenomXHM} is better, while at lower $q$'s the match with \texttt{BHPTNRSur2dq1e3} is better.
The effectualness comparisons in Fig.~\ref{fig:effectualness} (which gives a sensitivity of $0.9^3 \sim 0.7$) are better than the results in this appendix because here we are comparing the match of waveforms with the same exact parameters, while for the effectualness calculation we are recording the best match across all templates for each waveform sample.

\begin{figure}[t!]
    \centering
    \includegraphics[width=0.99\columnwidth]{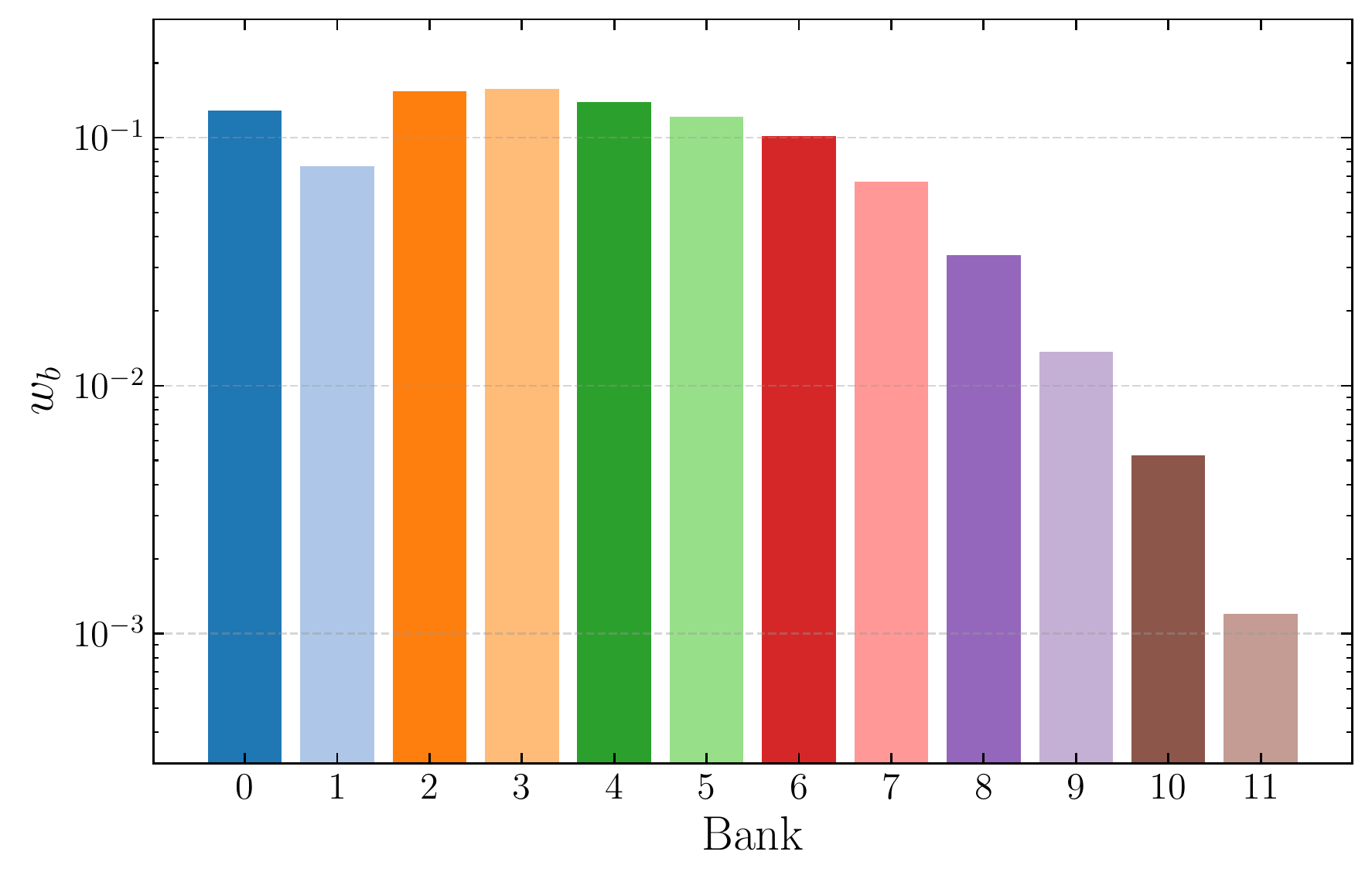}
\caption{
Our astrophysical prior for the fraction of events $w_b$ expected to be detected for each bank, as estimated with Eq.~\eqref{eq:bank_weights}. 
We also factored in the detector sensitivity, which is why the high-mass banks are down weighted as the signal progressively moves away from the sensitive frequencies in the detector.
If we expect to detect fewer events in a bank, the triggers we see in that bank would more likely be a false alarm.
Therefore, the ${\rm FAR}_b$ of each bank should be weighted by $1/w_b$ to obtain the ${\rm FAR}_{\rm overall}$ combined over all banks, see Eq.~\eqref{eq:overall_FAR}.
Note that the reported ${\rm FAR}_{\rm overall}$ of candidates are therefore dependent on the choice of the astrophysical prior, especially if the prior varies strongly across the search parameter space. 
}

    \label{fig:weight_banks} 
\end{figure}

\section{From bank specific to overall FAR}\label{app:weight_banks}

If we expect to detect the same number of GW events in all of the banks, then the ${\rm FAR}_{\rm overall}$ combined over all of the banks is just $N_{\rm banks}$ times the bank specific ${\rm FAR}_{\rm b}$ ($b \in \{0, 1, \dots, N_{\rm banks}\}$), because false alarms will occur $N_{\rm banks}$ times as often when all banks are considered.
However, even if we assume that the astrophysical rate-density of events within the intrinsic parameter coverage of the banks is the same across all banks, the sensitivity of each bank is different, causing the expected number of detected events to be different.
Therefore, to compute ${\rm FAR}_{\rm overall}$, we weight the bank-specific ${\rm FAR}_{\rm b}$ by a weight $w_b$
\begin{align}
    {\rm FAR}_{\rm overall} &= \sum_{b} \dfrac{{\rm FAR}_{b}}{w_b}, \label{eq:overall_FAR}\\
    w_b &\sim \sum_{i\in{\rm bank}\, b}\langle h_{22,i}\, | \,h_{22,i} \rangle^{3/2} \sin\iota_i, \label{eq:bank_weights}
\end{align}
with $h_{22,i}$ the $\ell m = 22$ mode of the $i$th waveform within a bank.
The summation is restricted to the waveforms within the bank $b$, not across different banks, and these waveforms are sampled from the astrophysical prior specified in Appendix~\ref{app:astro_search_prior}.
The square root of the inner product is the optimal SNR for that waveform and its cube is its sensitivity volume, while the $\sin\iota_i$ term accounts for the dependence of the $22$-mode sensitivity on the inclination for individual events.
The weights $w_b$ are approximately the sensitivity volume of the bank as a whole, and they are further normalized such that $\sum_b w_b = 1$, meaning that they are the fraction of detected events expected for each bank.
They impose the fact that less sensitive banks will have more false alarm candidates.
The weights $w_b$ for all banks are shown in Fig.~\ref{fig:weight_banks}.
We expect to detect a smaller fraction of events in the higher banks because higher mass events merge at lower frequencies, and the detector sensitivity is lower there.

\begin{table*}[ht!]
\centering
\setlength{\tabcolsep}{3pt}
\label{tab:gw_candidates_lvk}
\begin{tabular}{*{11}{c}}
\toprule
\multirow{2.5}{*}{Event name} & \multirow{2.5}{*}{$t_{\mathrm{gps}}$ [s]} & \multirow{2.5}{*}{Bank} & \multicolumn{4}{c}{Best-fit template parameters} & \multicolumn{2}{c}{IFAR [yr]} & \multirow{2.5}{*}{$\rho^2_\mathrm{H}$} & \multirow{2.5}{*}{$\rho^2_\mathrm{L}$} \\
\cmidrule(lr){4-7} \cmidrule(lr){8-9}
\rule[-3pt]{0pt}{0pt} &  &  & $m_1$ [$M_{\odot}$] & $1/q$ & $s_{1z}$ & $s_{2z}$ & bank & overall &  &  \\
\midrule
\multicolumn{11}{c}{\textbf{O3a}} \\
\addlinespace[0.5em]
\rowcolor{gray!15} GW190412\_053044 & 1239082262.10 & 0 & 90.4 & 19.4 & 0.70 & 0.41 & > 581 & 52.80 & 84.4 & 242.0 \\
 GW190707\_093326 & 1246527224.04 & 0 & 60.1 & 18.3 & 0.78 & -0.93 & > 581 & 44.68 & 60.0 & 96.3 \\
\rowcolor{gray!15} GW190728\_064510 & 1248331528.40 & 0 & 70.8 & 21.9 & 0.86 & -0.33 & > 581 & 44.68 & 58.1 & 103.7 \\
 GW190408\_181802 & 1238782700.24 & 1 & 86.9 & 18.3 & 0.22 & -0.82 & 290.4 & 38.72 & 89.4 & 94.7 \\
\rowcolor{gray!15} GW190828\_065509 & 1251010527.81 & 1 & 111.2 & 19.5 & 0.79 & -0.93 & 290.4 & 17.60 & 59.4 & 51.7 \\
 GW190513\_205428 & 1241816086.71 & 1 & 81.4 & 18.4 & -0.19 & 0.53 & 290.4 & 16.59 & 70.5 & 55.3 \\
\rowcolor{gray!15} GW190602\_175927 & 1243533585.07 & 3 & 125.7 & 18.0 & -0.24 & 0.20 & 52.8 & 13.51 & 45.6 & 107.1 \\
 GW190720\_000836 & 1247616534.57 & 0 & 62.6 & 18.7 & 0.77 & -0.05 & 580.8 & 8.67 & 39.7 & 55.3 \\
\rowcolor{gray!15} GW190929\_012149 & 1253755327.48 & 3 & 123.8 & 18.0 & -0.28 & 0.77 & 34.2 & 6.18 & 39.3 & 61.2 \\
 GW190503\_185404 & 1240944862.27 & 2 & 112.0 & 18.2 & 0.04 & -0.77 & 58.1 & 5.64 & 78.4 & 50.1 \\
\rowcolor{gray!15} GW190517\_055101 & 1242107479.81 & 3 & 106.1 & 34.8 & -0.94 & 0.50 & 11.0 & 1.73 & 46.5 & 62.2 \\
 GW190930\_133541 & 1253885759.10 & 0 & 58.5 & 18.1 & 0.79 & -0.78 & 22.3 & 1.72 & 37.7 & 52.0 \\
\rowcolor{gray!15} GW190725\_174728 & 1248112066.31 & 0 & 56.3 & 18.4 & 0.81 & -0.05 & 13.8 & 1.30 & 25.4 & 65.9 \\
 GW190706\_222641 & 1246487219.31 & 7 & 222.1 & 22.1 & -0.88 & 0.43 & 2.62 & 0.15 & 81.5 & 68.1 \\
\rowcolor{gray!15} GW190514\_065416 & 1241852074.84 & 3 & 113.8 & 18.4 & -0.93 & 0.02 & 0.44 & 0.08 & 38.7 & 25.1 \\
 GW190421\_213856 & 1239917954.23 & 2 & 78.9 & 18.4 & -0.48 & 0.36 & 0.32 & 0.06 & 76.8 & 31.9 \\
\rowcolor{gray!15} GW190926\_050336 & 1253509434.04 & 2 & 71.6 & 20.9 & -0.82 & 0.79 & 0.13 & 0.03 & 49.6 & 28.6 \\
 GW190916\_200658 & 1252699636.87 & 2 & 103.0 & 18.8 & -0.18 & -0.48 & 0.10 & 0.02 & 39.7 & 35.6 \\
\addlinespace[1.0em]
\multicolumn{11}{c}{\textbf{O3b}} \\
\addlinespace[0.5em]
 GW200316\_215756 & 1268431094.03 & 0 & 64.1 & 18.3 & 0.78 & -0.27 & 527.6 & 9.42 & 33.0 & 72.0 \\
\rowcolor{gray!15} GW191215\_223052 & 1260484270.28 & 2 & 149.0 & 18.4 & 0.73 & -0.02 & 15.1 & 1.27 & 28.6 & 55.2 \\
 GW191127\_050227 & 1258866165.52 & 2 & 77.6 & 21.2 & -0.44 & 0.62 & 10.6 & 0.97 & 55.5 & 48.2 \\
\rowcolor{gray!15} GW191105\_143521 & 1256999739.79 & 0 & 61.6 & 20.3 & 0.78 & -0.60 & 3.94 & 0.44 & 32.6 & 56.0 \\
 GW200219\_094415 & 1266140673.17 & 2 & 78.1 & 18.7 & -0.64 & -0.87 & 2.52 & 0.35 & 39.2 & 70.6 \\
\rowcolor{gray!15} GW200216\_220804 & 1265926102.85 & 2 & 95.7 & 21.9 & -0.27 & 0.58 & 2.16 & 0.32 & 49.5 & 38.0 \\
 GW200128\_022011 & 1264213229.86 & 5 & 189.1 & 28.6 & 0.12 & 0.70 & 2.64 & 0.31 & 32.3 & 42.0 \\
\rowcolor{gray!15} GW200208\_130117 & 1265202095.92 & 2 & 107.1 & 20.0 & -0.11 & -0.55 & 1.86 & 0.28 & 34.1 & 43.3 \\
 GW191230\_180458 & 1261764316.39 & 7 & 240.5 & 36.3 & -0.03 & -0.44 & 2.49 & 0.14 & 46.4 & 43.8 \\
\rowcolor{gray!15} GW191126\_115259 & 1258804397.52 & 0 & 80.1 & 23.9 & 0.83 & -0.22 & 0.05 & 0.01 & 30.2 & 45.4 \\
\bottomrule
\end{tabular}
\caption{Events in the GWTC-3 catalog recovered by our IMRI search pipeline.
The column headings are the same as that of Table~\ref{tab:gw_candidates}.
Note that none of these events have a mass ratio $q < 1/18$ after further analysis with Bayesian parameter estimation.
They are recovered by the listed templates in our pipeline simply because they roughly resemble them and are loud enough to trigger them. 
We only show the events recovered with $\mathrm{IFAR}_{\rm overall} \geq 0.01$ years.
}
\end{table*}

\section{GWTC-3 candidates}\label{app:lvk_cands}

In Table~\ref{tab:gw_candidates} of Sec.~\ref{sec:candidates}, we presented the candidates recovered by our search pipeline, but we only reported the candidates that do not correspond to events already reported in previous catalogs.
While our templates are constructed in the IMRI regime, they might still resemble more equal mass BBH waveforms and pick up those events that are loud enough.
In fact, we do recover many events previously reported in the GWTC-3 catalog~\cite{KAGRA:2021vkt}, and we report them in Table~\ref{tab:gw_candidates_lvk}.
Some of the louder events were recovered with a high IFAR.
Note, again, that the best-fit template parameters do not correspond to the actual parameters of the event, because all the events were found to have $q > 1/18$ after Bayesian parameter estimation.
Nonetheless, most of the events are recovered by templates close to the $q \sim 1/18$ upper edge of the parameter space, which is consistent with the fact that all the events have $q > 1/18$.

\section{Consistency of IFAR distribution with noise}\label{app:ifar_cum}

If no GW signals are present in the data stream, the number of candidates found with an IFAR greater than $\tau$ (i.e., a FAR less than $r \equiv 1/\tau$) follows a Poisson distribution $\mathrm{Pois}(1/\tau)$, which has expected value $1/\tau$.
In Sec.~\ref{sec:candidates}, disregarding the candidates that correspond to GWTC-3 events, we do not find any significant candidates.
In that case, the number of candidates with $\mathrm{IFAR} > \tau$ should be close to $1/\tau$ within the Poisson error.
In Fig.~\ref{fig:ifar_cum}, for the candidates found in O3a and O3b, we show that the cumulative distribution of $\mathrm{IFAR} > \tau$ is indeed consistent with $\mathrm{Pois}(1/\tau)$ for all $\tau$, verifying that the data stream consists purely of noise.

\begin{figure*}[t!]
    \centering
    \includegraphics[width=0.8\textwidth]{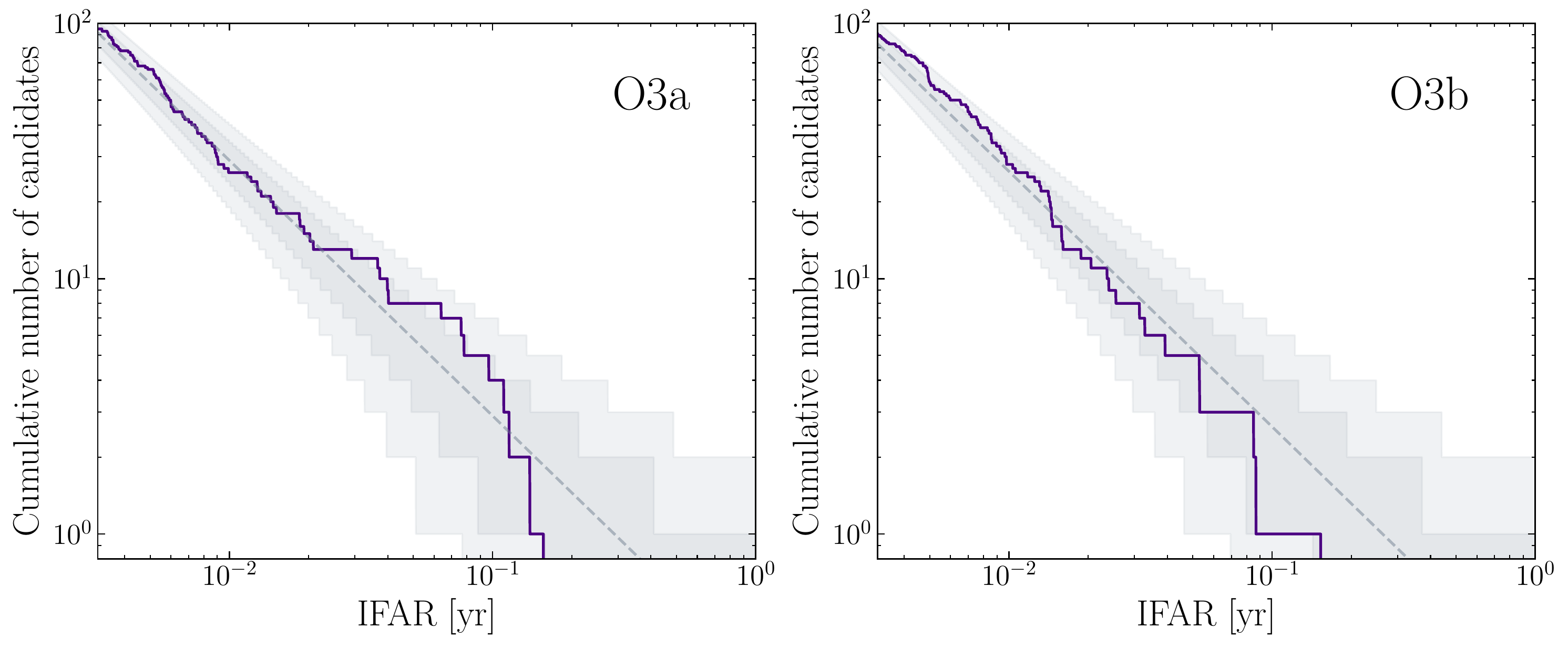}
    \caption{The distribution of overall IFAR for the recovered candidates in O3a (left) and O3b (right).
    The cumulative number of candidates above a given IFAR (solid lines) is close to the expected value (dashed line) and within the 1-$\sigma$ and 2-$\sigma$ confidence intervals (shaded regions) computed with the Poisson distribution, meaning that our candidates have IFARs that are consistent with noisy triggers.
    In this plot, we do not include the GWTC-3 candidates that we recovered.
    }
    \label{fig:ifar_cum} 
\end{figure*}

\clearpage

\bibliography{imri_template}

\end{document}